\documentclass[preprint,showpacs,preprintnumbers,amsmath,amssymb]{revtex4}%
\usepackage{graphicx}
\usepackage{dcolumn}
\usepackage{bm}
\usepackage{amsmath}
\usepackage{epstopdf}
\usepackage{amsfonts}
\usepackage{amssymb}%
\providecommand{\U}[1]{\protect\rule{.1in}{.1in}}
\newcommand{\be}{\begin{equation}}
\newcommand{\en}{\end{equation}}
\newcommand{\bea}{\begin{eqnarray}}
\newcommand{\ena}{\end{eqnarray}}
\begin{document}
\title{Impact of generalized dissipative coefficient on warm inflationary dynamics in the light of latest Planck data}
\author{Abdul Jawad}
\email{abduljawad@ciitlahore.edu.pk; jawadab181@yahoo.com}
\affiliation{Department of Mathematics, COMSATS Institute of Information Technology, Lahore-54000, Pakistan.}
\author{Shahzad Hussain}
\email{shahzad\_qau19@yahoo.com}
\affiliation{Department of Mathematics, COMSATS Institute of Information Technology, Lahore-54000, Pakistan.}
\author{Shamaila Rani}
\email{shamailatoor.math@yahoo.com; drshamailarani@ciitlahore.edu.pk}
\affiliation{Department of Mathematics, COMSATS Institute of Information Technology, Lahore-54000, Pakistan.}
\author{Nelson Videla}
\email{nelson.videla@pucv.cl}
\affiliation{Instituto de F\'isica, Pontificia Universidad Cat\'olica de Valpara\'iso.\\ Avda. Universidad 330, Curauma, Valpara\'iso, Chile.}

\date{\today}

\begin{abstract}
The warm inflation scenario in view of the
modified Chaplygin gas is studied. We consider 
the inflationary expansion is driven by a standard scalar field whose
decay ratio $\Gamma$ has a generic power law dependence with the scalar field $\phi$ and the 
temperature of the thermal bath $T$. By assuming an exponential power law dependence in the cosmic 
time for the scale factor $a(t)$, corresponding to the intermediate inflation model, we solve the background and perturbative dynamics considering that our model evolves according to
(i) weak dissipative regime and (ii) strong dissipative regime. Specifically, we find explicit expressions for the dissipative coefficient, scalar potential,  and the relevant inflationary observables
as the scalar power spectrum, scalar spectral index, and tensor-to-scalar ratio. The free parameters characterizing our model are constrained by considering the essential condition for warm inflation,  the conditions for the model evolves according to weak  or strong  dissipative regime, and the 2015 Planck results through the $n_s-r$ plane. 
\end{abstract}

\maketitle



\section{Introduction}

Inflation is the most acceptable paradigm that describes the physics of the very early universe. Besides of solving most of the shortcomings of the hot big-bang scenario, like the horizon, the flatness, and the
monopole problems \cite{R1,R106,R103,R104,R105,Linde:1983gd}, inflation also generates a mechanism to explain the large-scale structure (LSS) of the universe  \cite{R2,R202,R203,R204,R205}
and the origin of the anisotropies observed in the cosmic microwave background (CMB) radiation \cite{astro,astro2,astro202,Hinshaw:2012aka,Ade:2013zuv,Ade:2013uln,Ade:2015xua,Ade:2015lrj}, since primordial density perturbations may be sourced from quantum fluctuations of the inlaton scalar field during the inflationary expansion.

The standard cold inflation scenario is divided into two regimes: the slow-roll and reheating phases. In the slow-roll period the universe undergoes an accelerated expansion and all interactions
between the inflaton scalar field and other field degrees of freedom are typically neglected. Subsequently, a
reheating period \cite{Kofman:1994rk, Kofman:1997yn} is invoked to end the brief acceleration. After reheating, the universe is filled
with relativistic particles and thus the universe enters in the radiation big-bang epoch. For a modern review of reheating, see \cite{Amin:2014eta}. On the other hand, warm inflation is an alternative mechanism for having successful inflation. The warm inflation
scenario, as opposed to standard cold inflation, has the essential feature that a reheating phase is avoided at the end of the accelerated expansion due to the decay of the inflaton into radiation
and particles during the slow-roll phase \cite{warm1,warm2}. During warm inflation, the temperature of the universe does not drop dramatically and the universe can smoothly enter into the decelerated, radiation-dominated period, which is essential for a successful
big-bang nucleosynthesis. In the warm inflation scenario, dissipative effects are important during the accelerated expansion,
so that radiation production occurs concurrently with the accelerated expansion. The dissipative effect arises from a friction term $\Gamma$ which describes the processes of the scalar field dissipating into a thermal bath via its interaction with other field degrees of freedom. The effectiveness of warm inflation may be parametrized by the ratio $R\equiv \Gamma/3H$. The weak dissipative regime for warm inflation is for $R\ll 1$, while for $R\gg1$, it is the strong dissipative regime for warm inflation. Following Refs.\cite{Zhang:2009ge,BasteroGil:2012cm},
a general parametrization of the dissipative coefficient depending on both the temperature
of the thermal bath $T$ and the inflaton scalar field $\phi$ can be written as
\begin{equation}
\Gamma(T,\phi)=C_{\phi}\,\frac{T^{m}}{\phi^{m-1}}, \label{G}%
\end{equation}
where  the parameter $C_{\phi}$ is related   with  the dissipative
microscopic dynamics and the exponent $m$ is an integer, where the value of the power $m$ dependent on the specifics of the model construction for warm inflation and on the temperature regime of the thermal bath. Typically, it is found that $m=3$ (low temperature), $m=1$ (high temperature) or $m=0$ (constant dissipation). Additionally, thermal fluctuations during
the inflationary scenario may play a fundamental  role in
producing the primordial fluctuations \cite{62526,1126}. During the warm
inflationary scenario the density perturbations arise from
thermal fluctuations of the inflaton  and dominate over the
quantum ones. In this form,
an essential  condition for warm inflation to occur is the
existence of a radiation component with temperature $T>H$, since the thermal and quantum
fluctuations are proportional to $T$ and $H$,
respectively\cite{warm1,62526,1126,6252603,6252604,Taylor:2000ze}. When the universe heats
up and becomes radiation dominated, inflation ends and the
universe smoothly enters in the radiation
Big-Bang phase\cite{warm1}. For a comprehensive review of warm
inflation, see Ref. \cite{Berera:2008ar}.

The observational data from the luminosity-redshift of type Ia supernovae (SNIa), large
scale structure (LSS), and the cosmic microwave background (CMB) anisotropy spectrum,
have supported evidence that our universe has started recently a phase of accelerated expansion
\cite{astro,astro2,astro202,Hinshaw:2012aka,Ade:2013zuv,Ade:2013uln,Ade:2015xua,Ade:2015lrj,a1,a2,a3,a4}. The responsible for this acceleration of the late expansion is an exotic component having a negative pressure, usually known as dark energy (DE). Several models have been already proposed to be DE candidates, such as cosmological constant \cite{de1}, quintessence \cite{de2,de3,de4},
k-essence \cite{de5,de6,de7}, tachyon \cite{de8,de9,de10}, phantom \cite{de11,de12,de13}, Chaplygin gas \cite{de14}, holographic dark energy \cite{Li:2004rb}, among others in order to modify the matter sector of the gravitational action. Despite the plenty of models, the nature of the dark sector of the universe, i.e. dark energy and dark matter, is still unknown. There exists another way of understanding the observed universe in which dark matter and dark energy are described by a single unified component. Particularly,  the Chaplygin gas \cite{de14} achieves the unification of dark energy and dark matter. In this sense,  the Chaplygin gas behaves as a pressureless matter at the early times, and like a cosmological constant at late times. The original Chaplygin gas is characterized by an exotic equation of state with negative pressure 
\begin{equation}
p_{cg}=-\frac{B}{\rho_{cg}},\label{ocg}
\end{equation}
whit $B$ being a constant parameter. The original Chaplygin gas has been extended to the so-called generalized Chaplygin (GCG) gas with the following equation of state \cite{gcg}
\begin{equation}
p_{gcg}=-\frac{B}{\rho_{gcg}^{\lambda}},\label{ocg}
\end{equation}
with $0\leq \lambda \leq 1$. For the particular case $\lambda=1$, the original Chaplygin gas is recovered. The main motivation for studying this kind of model comes from string theory. The Chaplygin gas emerges as an
effective fluid associated with D-branes which may be obtained from the Born-Infeld action \cite{bi}. At background level, the GCG is able to describe the cosmological dynamics \cite{Makler:2002jv}, however the model presents serious issues at perturbative level \cite{Amendola:2003bz}. Thus, a modification to the GCG, resulting in the modified Chaplygin gas (MCG) with a equation of state given by \cite{Benaoum:2002zs}
\begin{equation}
p_{mcg}=A\rho_{mcg}-\frac{B}{\rho_{gcg}^{\lambda}},\label{mcg}
\end{equation}
where $A$, $B$, are constant parameters, with $0\leq \lambda\leq1$, is suitable to describe the evolution of the universe \cite{Lu:2008zzb,Debnath:2004cd} and it is also consistent with perturbative study \cite{SilvaeCosta:2007xy}.

As we have seen, the original and generalized Chaplygin gas models are
usually applied to explain the late time acceleration of our
universe as a possible candidate of dark energy. On the other hand, the modified Chaplygin
gas (MCG) is also a model that mimics the behavior
of matter at early-times and that of a cosmological constant at
late-times. Given the attractiveness of the MCG as a dark energy candidate, a
natural question to ask is: Can inflation be accommodated within the
MCG scenario? This is the question we wish to address in the present
work. However, we should emphasize that our inflationary model is
not presented as a more desirable alternative to the conventional
ones. Rather, we merely aim to establish the assumptions and
extrapolations required to obtain successful inflation in a
Chaplygin inspired model \cite{Bertolami:2006zg}.

Various authors have examined the warm inflation by considering
Chaplygin gas, standard  and tachyon scalar field models in
Einstein's General Relativity as well as in brane-world scenario with different
expressions for the dissipative coefficient \cite{22H}-\cite{mj7}. They found
the consistency of their results with observational data i.e,
BICEP$2$, WMAP $(7+9)$ and Planck data. Moreover, many authors have
investigated the warm inflation in various alternative as well as
modified theories of gravity \cite{bam1}-\cite{bam5}. Recently,
Herrera et al. \cite{18} studied the warm intermediate inflation in
the context of GCG in the weak and strong dissipative regimes by
assuming a generalized form of the dissipative coefficient under
slow-roll approximation. They found the constraints on the
parameters by considering the Planck 2015 data, together with the
essential condition for warm inflation $T > H$.

The main goal of the present work is to investigate the dynamics
of warm inflation driven by a standard scalar field in the
MCG scenario, with an inflaton decay rate  $\Gamma$ given by
the generalized expression (\ref{G}). By assuming an exponential power law dependence in the cosmic 
time for the scale factor $a(t)$, we solve the background and perturbative dynamics considering that our model evolves according to
(i) weak dissipative regime ($R \ll 1$) and (ii) according to strong dissipative regime ($R\gg 1$). The free parameters characterizing our model are constrained by considering the essential condition for warm inflation, $T>H$, the condition for the model evolves according to weak  or strong  dissipative regime, and the 2015 Planck results through the $n_s-r$ plane.

This paper is organized as follows: In the next section, we present
the basic setup of warm inflation in the MCG scenario. In sections
\ref{weak} and \ref{strong}, we solve the background and perturbative dynamics
when the model evolves according to weak and strong regimes, respectively. Specifically, in each section, we find explicit expressions for the dissipative coefficient, scalar potential,  and the relevant inflationary observables
as the scalar power spectrum, scalar spectral index, and tensor-to-scalar ratio. Finally, section \ref{conclu} summarizes our finding and exhibits our conclusions. We have chosen units such that $c=\hbar=1$.

\section{Modified Chaplygin Gas Inspired Inflation}\label{mgcbs}

In this section, we introduce the basic setup of warm inflation in
MCG scenario with a generalized expression for the
inflaton decay rate $\Gamma$. As it was mencioned at the introduction, the exotic equation of state
of MCG is given by
\begin{equation}\label{mcg1}
p_{mcg}=A\rho_{mcg}-\frac{B}{\rho_{mcg}^\lambda},
\end{equation}
where $A$ and $B$ are constant parameters with $0\leq \lambda\leq1$. $p_{mcg}$ and
${\rho}_{mcg}$ are the pressure and energy density of
MCG, respectively. The energy density of MCG as function of the scale factor $a$ can be obtained with the help of
the stress-energy conservation law, yielding
\begin{equation}\label{mcg2}
\rho_{mcg}=\left[\frac{B}{1+A}+\frac{C}{a^{3(1+\lambda)(1+A)}}\right]^\frac{1}{1+\lambda}=\rho_{mcg0}\left[B_s+\frac{1-B_s}{a^{3(1+\lambda)(1+A)}}\right]^\frac{1}{1+\lambda},
\end{equation}
where $B_s=\frac{B}{1+A}\frac{1}{\rho^{1+\lambda}_{mcg0}}$, $C$ is a positive integration constant. From the solution given by Eq.(\ref{mcg2}), the energy density of the MCG is characterized by three
parameters, $B_s$ (or equivalently $B$), $A$, and $\lambda$. Particularly, in \cite{Paul:2014kza} by using a joint analysis of several tests at background as well as perturbative level, as the differential age of old galaxies, given by $H (z)$, Baryonic acoustic oscillations (BAO) peak parameter, CMB shift parameter, SN Ia data, and growth index, the values for the best-fit (with $\chi^2/{d.o.f}\sim 1\textup{.}0296$) are given by  $B_s=0\textup{.}8252$, $A=0\textup{.}0046$, and $\lambda=0\textup{.}1905$.

As was mentioned in the introduction, in order to obtain successful inflation in a Chaplygin like inspired model, some
assumptions and extrapolations are required. Following \cite{Bertolami:2006zg}, we identify the energy density of matter $\rho_m$ with the contribution of the energy density associated to the standard scalar field $\rho_{\phi}$ through an extrapolation of Eq.(\ref{mcg2}), yielding

\begin{eqnarray}\label{mcg3}
\left[\frac{B}{1+A}+\rho^{(1+\lambda)(1+A)}_m\right]^{\frac{1}{1+\lambda}}
\rightarrow \left[\frac{B}{1+A}+\rho_{\phi}^{(1+\lambda)(1+A)}\right]^{\frac{1}{1+\lambda}}.
\end{eqnarray}

In this sense, we will not consider Eq.(\ref{mcg3}) as a consequence of Eq.(\ref{mcg2}), but a non-covariant modification of gravity instead, resulting in a modifed Friedmann equation, as it was pointed up in \cite{Barreiro:2004bd}.

In this scenario, we consider a spatially flat universe which contains a self-
interacting inflation field $\phi$ and a radiation field, then we write
down a modified Friedmann equation of the form
\begin{equation}\label{f1}
H^{2}=\frac{\kappa}{3}\bigg(\left[\frac{B}{1+A}+\rho_{\phi}^{(1+\lambda)(1+A)}\right]^{\frac{1}{1+\lambda}}+\rho_{\gamma}\bigg),
\end{equation}
where $\kappa=8{\pi}G$ and $H$ is the Hubble rate defined
as $H=\dot{a}/a$. 

We recall that Friedmann equation (\ref{f1}) constitutes a non-covariant modification of gravity. However, as it was pointed up in Ref.\cite{Bertolami:2006zg},
it may assumed that the effect
giving rise to Eq.(\ref{f1}) preserves diffeomorphism invariance in (3+1)
dimensions, whence total stress-energy conservation follows. In this way, for our analysis, the second Friedmann equation is no longer requiered. 

By coupling the inflaton field to a radiation fluid, the conservation equations for each individual component are given by \cite{warm1,warm2}
\begin{equation}\label{C1}
\dot{\rho}_{\phi}+3H({\rho}_{\phi}+P_{\phi})=-\Gamma\dot{\phi},
~~\Longrightarrow~~
\ddot{\phi}+3H\dot{\phi}+V^{\prime}=-\Gamma\dot{\phi},
\end{equation}
and
\begin{equation}\label{C2}
{\dot{\rho}}_{\gamma}+4H{\rho}_\gamma=\Gamma{\dot{\phi}}^2,
\end{equation}
where $\rho_{\phi}=\frac{\dot{\phi}^2}{2}+V(\phi)$ and $P_{\phi}=\frac{\dot{\phi}^2}{2}-V(\phi)$ correspond to the energy density and pressure associated with the
standard scalar field, respectively, and $V(\phi)$ is the inflaton's potential. On the other hand, $\Gamma$ represents the inflaton decay rate or dissipative coefficient, which is
responsible for the process of decay of the scalar field into
radiation during the inflationary expansion. This decay rate can be realized
as a constant or can be a function of scalar field or
temperature or both, i.e., $\Gamma(T,\phi)$. From first principles in quantum field theory this decay ratio $\Gamma$
has been already computed. A generalized form of
$\Gamma$ is given by \cite{Zhang:2009ge,BasteroGil:2012cm}
\begin{equation}\label{D1}
\Gamma(T,\phi)=C_\phi\frac{T^m}{\phi^{m-1}}.
\end{equation}
In literature, several cases have been studied for the different
values of $m$, in special case $m=1$, i.e. $\Gamma\propto T$
represent high temperature SUSY case, for the value $m=0$ i.e.
$\Gamma\propto \phi$ corresponds to an exponentially decaying
propagator in the high temperature SUSY model, for $m=-1$ i.e.
$\Gamma\propto \frac{\phi^{2}}{T}$, with non-SUSY
case.

Considering that during warm inflation the energy
density associated  of radiation field $\rho_{\phi}\gg\rho_{\gamma}$ is
subdominat with respect to energy density of the scalar field \cite{warm1,warm2,62526,1126,6252603,6252604,Taylor:2000ze},
then Eq.(\ref{f1}) becomes
\begin{equation}\label{f2}
H^2\approx\frac{\kappa}{3}\bigg(\left[\frac{B}{1+A}+\rho_{\phi}^{(1+\lambda)(1+A)}\right]^{\frac{1}{1+\lambda}}\bigg)=\frac{\kappa}{3}\bigg(\left[\frac{B}{1+A}+\left(\frac{\dot{\phi}^2}{2}+V(\phi)\right)^{(1+\lambda)(1+A)}\right]^{\frac{1}{1+\lambda}}\bigg).
\end{equation}
By combining Eqs. (\ref{C1}) and (\ref{f2}), we obtain the square velocity of the inflaton field
\begin{equation}\label{sc1}
\dot{\phi}^2=\frac{2(-\dot{H})}{\kappa(1+A)(1+R)}\bigg(\frac{3H^2}{\kappa}\bigg)^\frac{-A}
{1+A}\left[1-\frac{B}{1+A}\bigg(\frac{3H^2}{\kappa}\bigg)^{-(1+\lambda)}\right]^{-\frac{A+\lambda(1+A)}
{(1+A)(1+\lambda)}}.
\end{equation}
In this equation, we have introduced a new parameter $R$ defined by
\begin{equation}\label{10}
R\equiv \frac{\Gamma}{3H}.
\end{equation}
This parameter measures the relative strength of thermal damping
compared to the expansion damping. In warm inflation, two possible
regimes can be described through $R$, i.e., weak dissipative regime
in which $R \ll 1$ and Hubble damping is still the dominant term in
this case. The second is strong dissipative regime which can be
defined as $R \gg 1$ and $\Gamma$ controls the damped evolution of the
inflation field.

By also assumming that ${\dot{\rho}}_\gamma\ll4H{\rho}_\gamma$, i.e.,
the radiation production is quasi-stable \cite{warm1,warm2,62526,1126,6252603,6252604,Taylor:2000ze},
Eqs.(\ref{C2}) and (\ref{sc1}) lead to the relation for
$\rho_\gamma$ as follows
\begin{eqnarray}\nonumber
\rho_\gamma&=&\frac{\Gamma\dot{\phi}^2}{4H}=\frac{\Gamma(-\dot{H})}
{2\kappa H(1+A)(1+R)}\bigg(\frac{3H^2}{\kappa}\bigg)^\frac{-A}{1+A}\left[1-\frac{B}{1+A}\bigg(\frac{3H^2}{\kappa}\bigg)^
{-(1+\lambda)}\right]^{-\frac{A+\lambda(1+A)}{(1+A)(1+\lambda)}},
\\\label{11}
\end{eqnarray}
In addition, the thermalized energy density of radiation field can
be written as $\rho_{\gamma}=C_\gamma T^{4}$, where 
$C_\gamma=\pi^{2}g_{*}/30$, and $g_{*}$ denotes the number of
relativistic degrees of freedom. In particular, for the Minimal Supersymmetric
Standard Model (MSSM), we have that $g_{*}=228.75$ and $C_\gamma\simeq 70$
\cite{BasteroGil:2012cm}. We can get the temperature of thermal bath from
Eq.(\ref{11}) as follows
\begin{equation}\label{12}
T=\left[\frac{\Gamma(-\dot{H})}{2\kappa {C_\gamma}(1+A)(1+R)}\right]^{1/4}
\bigg(\frac{3H^2}{\kappa}\bigg)^\frac{-A}{4(1+A)}\left[1-\frac{B}{1+A}\bigg(\frac{3H^2}{\kappa}\bigg)^
{-(1+\lambda)}\right]^{-\frac{A+\lambda(1+A)}{4(1+A)(1\lambda)}}.
\end{equation}
By considering Eqs.(\ref{f2}), (\ref{sc1}) and (\ref{12}), the
the inflaton's potential may be expressed as follows
\begin{eqnarray}\nonumber
V&=&\left[\bigg(\frac{3H^2}{\kappa}\bigg)-\frac{B}{1+A}\right]^\frac{1}{(1+A)(1+\lambda)}+\frac{\dot{H}}
{\kappa(1+A)(1+R)}\bigg(\frac{3H^2}{k}\bigg)^\frac{-A}{1+A}
\\\label{13}&\times&\left[1-\frac{B}{1+A}\bigg(\frac{3H^2}{\kappa}\bigg)^{-(1+\lambda)}\right]
^{-\frac{A+\lambda(1+A)}{(1+A)(1+\lambda)}}.
\end{eqnarray}
Similarly, by using Eqs.(\ref{D1}) and (\ref{12}), the dissipative
coefficient may be written as
\begin{eqnarray}\nonumber
\Gamma^\frac{4-m}{4}&=&{C_\phi}{\phi^{1-m}}
\left[\frac{(-\dot{H})}{2\kappa C_{\gamma}H(1+A)(1+R)}\right]^{m/4}\bigg(\frac{3H^2}{\kappa}\bigg)
^\frac{-mA}{4(1+A)}\bigg[1-\frac{B}{1+A}\\\label{14}&\times&\bigg(\frac{3H^2}{\kappa}\bigg)
^{-(1+\lambda)}\bigg]^{-\frac{m(A+\lambda(1+A))}{4(1+A)(1+\lambda)}}.
\end{eqnarray}

In the next two sections, we will explore the
inflationary dynamics at background as well as perturbative level when
our model evolves according to (i) weak dissipative regime and (ii) strong dissipative
regime, respectively. As an extra input, we assume an exponential
power-law dependence in cosmic time for the scale factor $a(t)$, given by the intermediate inflation model. Exact solutions 
in the context of inflation can be found from an exponential potential, obtaining
a solution for the scale factor give by $a(t)\sim t^p,~p>1$, termed as power-law
inflation \cite{Lucchin:1984yf}. On the other hand, by considering a constant scalar potential \cite{R1}, we obtain
an exponential solution for the scale factor $a(t) \sim \exp{H_0t}$, known as de-sitter expansion. For
an inverse power-law potential, the intermediate inflation model is found as an exact solution, for which
the scale factor expands faster than power-law expansion but slower
than de-sitter inflation. The scale factor $a(t)$ for
intermediate inflationary model is given by \cite{Barrow:1990vx}
\begin{equation}\label{3}
a(t)=\exp{[\alpha t^f]},
\end{equation}
where $\alpha >0$ and $0<f<1$. This model was in the beginning formulated as an exact solution
to the background equations, nevertheless this model may be studied under the slow-roll approximation together with the cosmological perturbations \cite{Barrow:1993zq,Rendall:2005if,Barrow:2006dh,Barrow:2014fsa}.

\section{The Weak Dissipative Regime}\label{weak}

Assuming that our model evolves according to the weak dissipative regime, i.e., $R\ll1$ (or
$\Gamma \ll 3H$), the scalar field $\phi$ as function of cosmic time may be
found by using Eqs.(\ref{sc1}) and (\ref{3}), yielding
\begin{equation}\label{15}
\phi(t)-{\phi_0}=\frac{M[t]}{S},
\end{equation}
where $\phi_0$ is a constant of integration, and the constant $S$ is
given by
\begin{equation*}
S=\frac{A(2-f)+f}{\sqrt{2(1-f)(1+A)}}\left(\frac{1}{\alpha f}\right)^{\frac{1-A}{2(1+A)}}\left(\frac{\kappa}{3}\right)^{\frac{1-A}{2(1+A)}}
\end{equation*}
while $M[t]$ is function of cosmic time taking the following
form
\begin{eqnarray}\nonumber
M[t]&=& t^{\frac{f+2A-Af}{2(1+A)}}\,_2F_1\bigg[\frac{A(2-f)+f}{4(1+\lambda)(1+A)(1-f)},
\frac{A+\lambda (1+A)}{2(1+A)(1+\lambda)},
\\\nonumber&&1+\frac{A(2-f)+f}{4(1+\lambda)(1+A)(1-f)},
\,\frac{B}{1+A}\,3^{-(1+\lambda)}t^{2(1-f)(1+\lambda)}\bigg(\frac{\kappa}{f^2{\alpha}^2}\bigg)^{1+\lambda}\bigg],
\end{eqnarray}
here $_2F_1$ denotes the  hypergeometric function \cite{Veberic:2012ax}
Under the slow-roll approximation, in which
${\dot{\phi}}^2/2<V(\phi)$, from Eq.(\ref{13}), the scalar
potential as a function of scalar field can be written as
\begin{equation}\label{16}
V(\phi)\approx\left[\bigg(\frac{3\alpha^2f^2}{\kappa(M^{-1}[S\phi])
^{2(1-f)}}\bigg)^{(1+\lambda)}-\frac{B}{1+A}\right]^\frac{1}{(1+A)(1+\lambda)}.
\end{equation}
In the similar way, we can obtain the dissipative coefficient in
terms of scalar field as
\begin{eqnarray}\nonumber
\Gamma(\phi)&=&\left[\frac{1-f}{2\kappa{C_\gamma}(1+A)(M^{-1}[S{\phi}])}\right]
^\frac{m}{4-m}\bigg(\frac{3\alpha^2f^2}{\kappa(M^{-1}
[S{\phi}])^{2(1-f)}}\bigg)^\frac{-m A}{(1+A)(4-m)}
\\\label{17}&\times&{C_{\phi}^{\frac{4}{4-m}}}{\phi}
^\frac{4(1-m)}{4-m)}\left[1-\frac{B}{1+A}\bigg(\frac{\kappa(M^{-1}[S{\phi}])^
{2(1-f)}}{3\alpha^2f^2}\bigg)^{(1+\lambda)}\right]
^{-\frac{m(A+\lambda(1+A))}{(1+A)(1+\lambda)(4-m)}}.
\end{eqnarray}
The number of $e$-folds, $N$, between two different values of cosmic
time, $t_1$ and $t_2$, or equivalently, between two values of the scalar field,
$\phi_1$ and $\phi_2$ is defined as follows
\begin{equation}\label{18}
N=\int_{t_1}^{t_2}Hdt=\alpha(t_{2}^{f}-t_{1}^{f})=
\alpha\bigg((M^{-1}[S{\phi}_2])^f-(M^{-1}[S{\phi}_1])^f\bigg).
\end{equation}

Since we are dealing with the scale factor $a(t)$, it is straightforward to use the slow-roll parameters
\begin{eqnarray}
&&\epsilon=-\frac{\dot{H}}{H^2},
\end{eqnarray}
and
\begin{eqnarray}
&&\eta=-\frac{\ddot{H}}{H\dot{H}}.
\end{eqnarray}

In the intermediate inflation model, the slow-roll parameters $\epsilon$ and $\eta$ decrease
as the field rolls down the potential, then there is no natural exit from the model \cite{Barrow:2006dh}. However, from the definition of the parameter $\epsilon$, we may obtain the value of
the scalar field for inflationary scenario at early stage
($\epsilon=1$) \cite{Barrow:2006dh}, giving
\begin{equation}\label{19}
\phi_1=\frac{1}{S}M\bigg[\bigg(\frac{1-f}{\alpha f}\bigg)^{1/f}\bigg].
\end{equation}

In this way, we may evaluate the inflationary observables at $N$ $e$-folds which have passed since the beginning of the inflationary
period.

In the following, we will study the scalar and tensor perturbations for
our warm inflation model in the MCG scenario, considering that it evolves 
according to the weak regime. For the case of the scalar perturbations, the amplitude could be 
stated as $\mathcal{P}_{\mathcal{R}}^{1/2}=\frac{H}{\dot{\phi}}\delta \phi$ \cite{Lyth:2009zz}.
Additionally, in the warm inflation scenario, a thermalized radiation component is present with $T > H$, then the inflaton fluctuations $\delta \phi$ are predominantly thermal instead quantum. Particularly, for the weak dissipation regime, the amplitude of the scalar field fluctuation was found to be $\delta \phi^2\simeq H T$ \cite{62526}. Then,
power spectrum of the scalar perturbations can be obtained by
utilizing Eqs.(\ref{sc1}), (\ref{12}) and (\ref{14})
\begin{eqnarray}\nonumber
\mathcal{P}_\mathcal{R}&=&\frac{\sqrt{3}}{4}\kappa(1+A)\bigg(\frac{3}{\kappa}\bigg)
^\frac{(3-m)A}{(1+A)(4-m)}\left[\frac{C_{\phi}}{2\kappa C_{\gamma}(1+A)}\right]
^\frac{1}{4-m}\phi^\frac{1-m}{4-m}(-\dot{H})^\frac{m-3}{4-m}
\\\label{20}&\times&(H)^\frac{(11-3m)(1+A)+2A(3-m)}{(1+A)(4-m)}
\left[1-\frac{B}{1+A}\bigg(\frac{3H^2}{\kappa}\bigg)^{-(1+\lambda)}\right]^\frac{(3-m)[A+\lambda(1+A)]}
{(4-m)(1+A)(1+\lambda)}.
\end{eqnarray}
The power spectrum of scalar perturbations in terms of scalar field
can also be written as
\begin{eqnarray}\nonumber
\mathcal{P}_\mathcal{R}&=&\delta_1\phi^\frac{1-m}{4-m}(M^{-1}[S\phi])
^\frac{(f-2)(m-3)(1+A)-(1-f)[(11-3m)(1+A)+2A(3-m)]}{(1+A)(4-m)}
\bigg[1-\frac{B}{1+A}\\\label{21}&\times&\bigg(\frac{\kappa(M^{-1}[S\phi])^
{2(1-f)}}{3\alpha^2f^2}\bigg)^{(1+\lambda)}\bigg]
^\frac{(3-m)[A+\lambda(1+A)]}{(4-m)(1+\lambda)(1+A)},
\end{eqnarray}
where $\delta_1$ is new constant which is given by
\begin{eqnarray}\nonumber
\delta_1=\frac{\sqrt{3}}{4}\kappa(1+A)\left[\frac{C_\phi}{2kC_\gamma(1+A)}\right]^\frac{1}{4-m}
(1-f)^\frac{m-3}{4-m}({\alpha}f)
^\frac{[(11-3m)(1+A)+2A(3-m)]-(3-m)(1+A)}{(1+A)(4-m)}
\end{eqnarray}
The power spectrum may also be written as a function of the number
of $e$-folds as follows
\begin{eqnarray}\nonumber
\mathcal{P}_\mathcal{R}(N)&=&\delta_2(M(J[N]))^\frac{1-m}{4-m}(J[N])
^\frac{(f-2)(m-3)(1+A)-(1-f)[(11-3m)(1+A)+2A(3-m)]}{(1+A)(4-m)}
\\\label{22}&\times&\left[1-\frac{B}{1+A}\bigg(\frac{\kappa(J[N])^{2(1-f)}}{3\alpha^2f^2}\bigg)
^{(1+\lambda)}\right]^\frac{(3-m)[A+\lambda(1+A)]}{(4-m)(1+A)(1+\lambda)}.
\end{eqnarray}
Where the constant $\delta_2$ is defined as
$\delta_2=\delta_1S^\frac{m-1}{4-m}$ and $J[N]$ is defined as
$J[N]=[\frac{1+f(N-1)}{Af}]^\frac{1}{f}$. Additionally, the scalar
spectral index $n_s$, defined by
$n_s-1=\frac{d\ln{\mathcal{P}_\mathcal{R}}}{d{\ln}k}$, and by using Eqs.
(\ref{15}) and (\ref{22}), takes the form as follows
\begin{eqnarray}\nonumber
n_s&=&1+\frac{(f-2)(m-3)(1+A)-(1-f)[(11-3m)(1+A)+2A(3-m)]} {\alpha 
f(1+A)(4-m)(M^{-1}[S\phi])^f}\\\label{23}&&+n_2+n_3,
\end{eqnarray}
where $n_2$ and $n_3$ are given by
\begin{eqnarray}\nonumber
n_2&=&\bigg(\frac{1-m}{4-m}\bigg)\sqrt{\frac{2(1-f)}{\kappa \alpha f(1+A)}}
\bigg(\frac{3{\alpha}^2f^2}{\kappa}\bigg)^{-\frac{A}{2(1+A)}}
\bigg(\frac{1}{\phi}\bigg)\bigg(M^{-1}[S\phi]\bigg)^{\frac{2A(1-f)-f(1+A)}{2(1+A)}}\\\nonumber&\times&
\left[1-\frac{B}{1+A}\bigg(\frac{\kappa\left(M^{-1}[S\phi]\right)^{2(1-f)}}{3{\alpha}^2f^2}\bigg)^{(1+\lambda)}\right]^{-\frac{A+\lambda(1+A)}{2(1+\lambda)(1+A)}},
\end{eqnarray}
and
\begin{eqnarray}\nonumber
n_3&=&-2\frac{B}{1+A}\bigg(\frac{3-m}{4-m}\bigg)\bigg(\frac{A+\lambda(1+A)}{(1+A)}\bigg)\frac{(1-f)(\kappa/3)
^{1+\lambda}}{(\alpha f)^{3+2\lambda}}\left(M^{-1}[S\phi]\right)
^{2-3f+2{\lambda}(1-f)}\\\nonumber&\times&\bigg[1-\frac{B}{1+A}\bigg(\frac{\kappa (M^{-1}[S\phi])
^{2(1-f)}}{3\alpha^2f^2}\bigg)^{(1+\lambda)}\bigg]^{-1}.
\end{eqnarray}
By using Eqs.(\ref{18}) and (\ref{19}), the scalar spectral index
may also be written in terms of number of $e$-folds $N$
\begin{eqnarray}\nonumber
n_s&=&1+\frac{(f-2)(m-3)(1+A)-(1-f)[(11-3m)(1+A)+2A(3-m)]}
{(4-m)(1+A)[1+f(N-1)]}\\\label{24}&&+n_2+n_3,
\end{eqnarray}
where $n_2$ and $n_3$ are defined as
\begin{eqnarray}\nonumber
n_2&=&S\bigg(\frac{1-m}{4-m}\bigg)\sqrt{\frac{2(1-f)}{\kappa \alpha f(1+A)}}
\bigg(\frac{3{\alpha}^2f^2}{\kappa }\bigg)^{-\frac{A}{2(1+A)}}\frac{(J[N])^{\frac{2A(1-f)-f(1+A)}{2(1+A)}}}{M(J[N])}
\\\nonumber&\times&\bigg[1-\frac{B}{1+A}\bigg(\frac{\kappa (J[N])^{2(1-f)}}{3{\alpha}^2f^2}\bigg)^{(1+\lambda)}\bigg]
^{-\frac{A+\lambda(1+A)}{2(1+\lambda)(1+A)}},
\end{eqnarray}
and
\begin{eqnarray}\nonumber
n_3&=&2\frac{B}{1+A}\bigg(\frac{3-m}{4-m}\bigg)\bigg(\frac{A+\lambda(1+A)}{(1+A)}\bigg)\frac{(1-f)(\kappa /3)
^{1+\lambda}}{(\alpha f)^{3+2\lambda}}(J[N]))^{2-3f+2{\lambda}(1-f)}
\\\nonumber&\times&\bigg[1-\frac{B}{1+A}\bigg(\frac{\kappa (J[N])^{2(1-f)}}{3\alpha^2f^2}\bigg)^{(1+\lambda)}\bigg]^{-1}.
\end{eqnarray}

Regarding tensor perturbations, these do not couple to the thermal background, so gravitational waves are only generated by quantum fluctuations, as in standard inflation \cite{Taylor:2000ze}
\begin{equation}\label{25}
\mathcal{P}_g=8\kappa \bigg(\frac{H}{2\pi}\bigg)^2.
\end{equation}
Having the tensor power spectrum, we may compute the tensor-to-scalar ratio
$r=\mathcal{P}_g/\mathcal{P}_\mathcal{R}$, yielding
follow
\begin{eqnarray}\nonumber
r(\phi)&=&\bigg(M^{-1}[S{\phi}]\bigg)
^\frac{(1-f)[(11-3m)(1+A)+2A(3-m)]-(f-2)(m-3)(1+A)-2(1-f)(1+A)(4-m)}
{(1+A)(4-m)}\\\label{26}&\times&\frac{2\kappa \alpha^2f^2}{{\pi}^2\delta_1}{\phi}^\frac{m-1}{4-m}
\bigg[1-\frac{B}{1+A}\bigg(\frac{\kappa (M^{-1}[S\phi])
^{2(1-f)}}{3\alpha^2f^2}\bigg)^{(1+\lambda)}\bigg]
^\frac{(m-3)[A+\lambda(1+A)]}{(4-m)(1+A)(1+\lambda)}.
\end{eqnarray}

\begin{figure}[th]
{{\hspace{0cm}\includegraphics[width=3.3in,angle=0,clip=true]{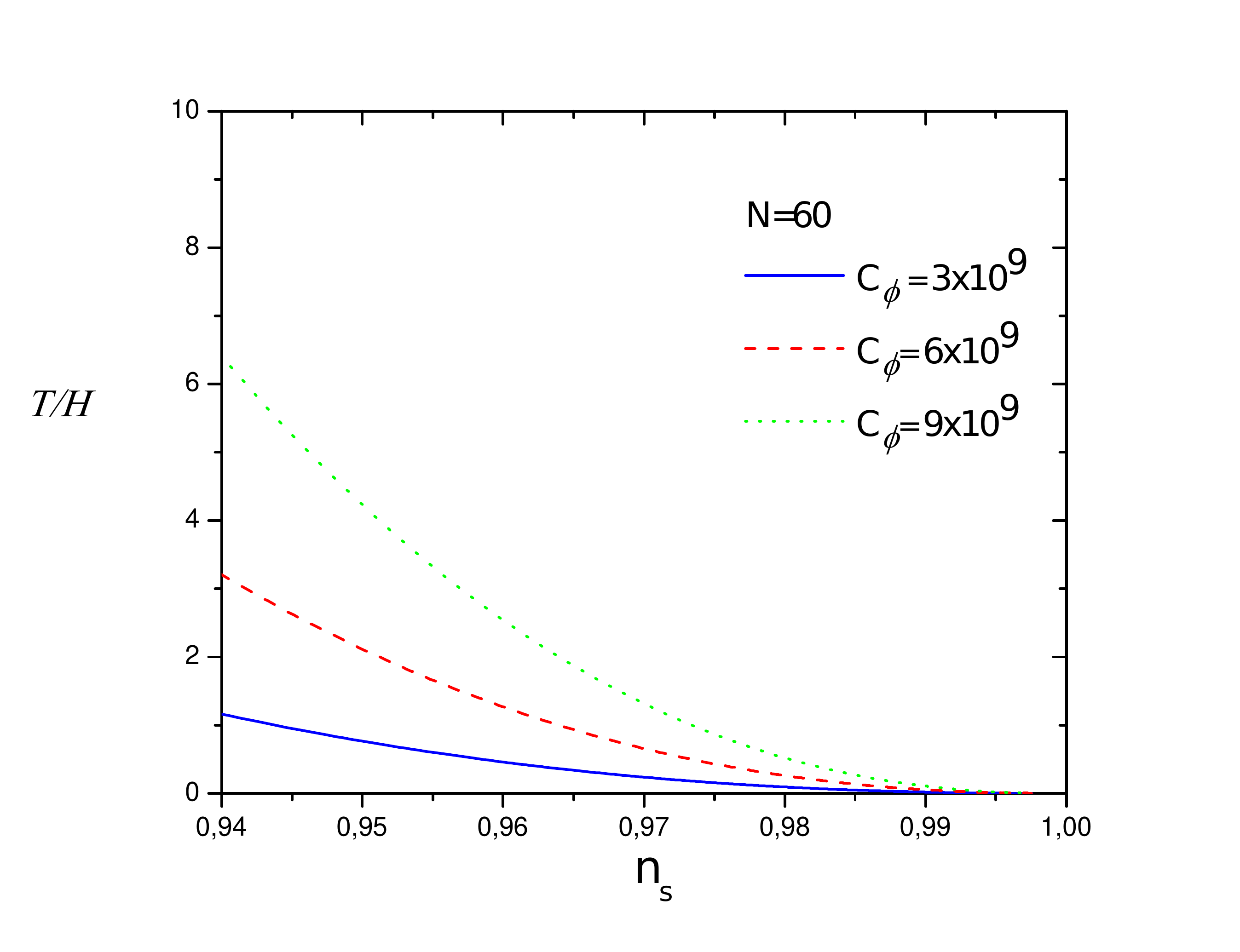}}
\hspace{-0.7 cm}}
{\includegraphics[width=3.3in,angle=0,clip=true]{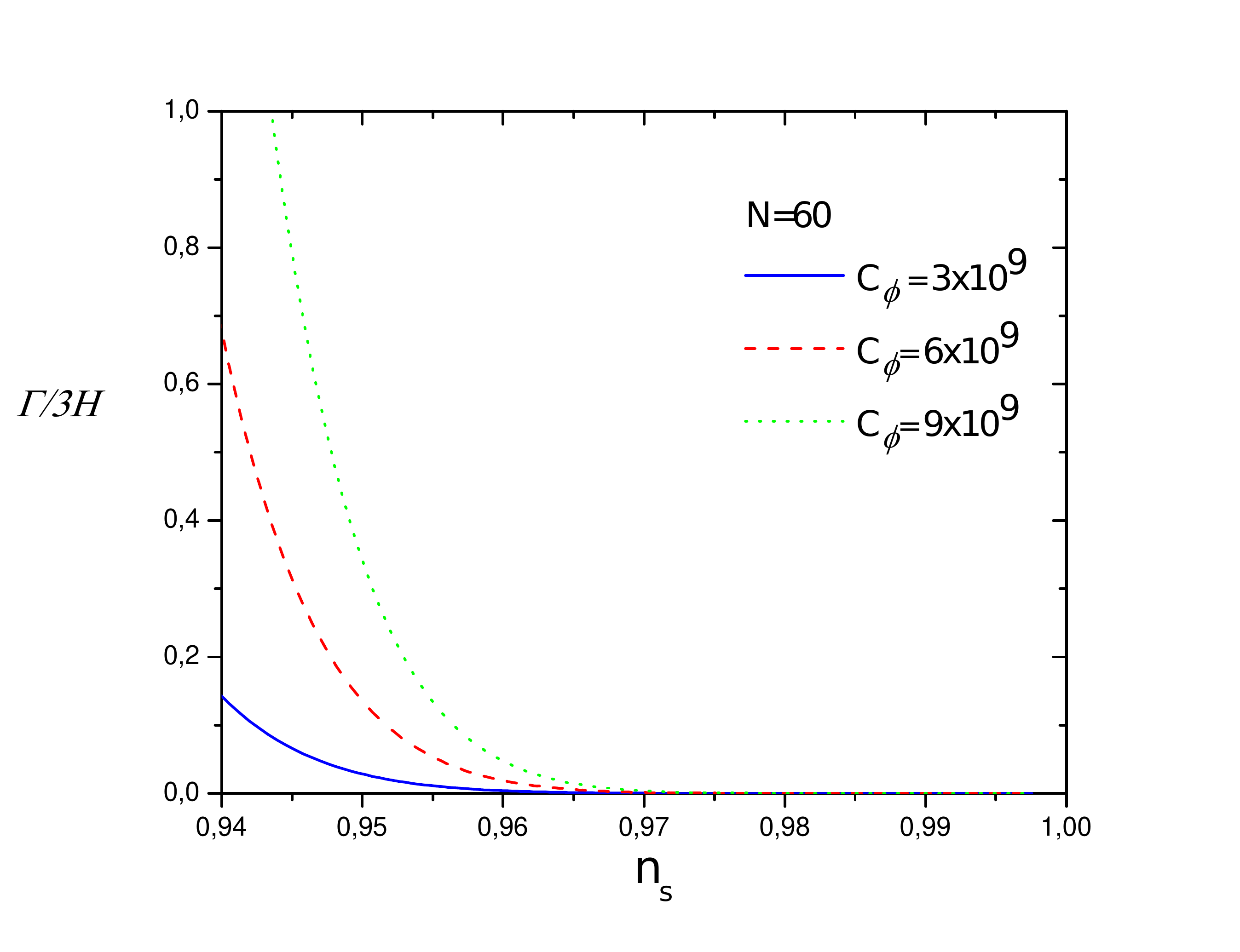}}
{\vspace{-0.5 cm}\caption{ Plots of $T/H$ as function of
the scalar spectral index $n_s$ (upper left) and $\Gamma/3H$ as function of
the scalar spectral index $n_s$ (upper right) for $N=60$. For both plots we have
considered three different values of the parameter $C_\phi$ for the
special case $m=3$, i.e., $\Gamma\propto T^3/\phi^2$, assuming 
the model evolves according to the weak dissipative regime. In both panels, the dotted, dashed, and
solid lines correspond to the pairs ($\alpha=0\textup{.}0062$, $f=0\textup{.}4703$),
($\alpha=0\textup{.}0043$, $f=0\textup{.}4702$), and ($\alpha=0\textup{.}0036$, $f=0\textup{.}4701$),
respectively. Lower left and lower right panels show $\Gamma/3H$ as function of the scalar spectral index $n_s$ for $N=55$ and $N=70$, respectively. In these plots we have used the values
$C_\gamma=70$, $A=0\textup{.}0046$, $B=0\textup{.}8289$, $\lambda=0\textup{.}1905$, and $\kappa=1$
 \label{fig1}}}
\end{figure}

Similarly, in terms of number of $e$-folds $N$, the tensor-to-scalar
ratio becomes
\begin{eqnarray}\nonumber
r(N)&=&(J[N]))^\frac{(1-f)[(11-3m)(1+A)+2A(3-m)]-
(f-2)(m-3)(1+A)-2(1-f)(1+A)(4-m)}{(1+A)(4-m)}
\\\label{27}&\times&\frac{2\kappa \alpha^2f^2}{{\pi}^2\delta_2}(M(J[N]))^\frac{m-1}{4-m}\bigg[1-\frac{B}{1+A}\bigg(\frac{\kappa (J[N]))^{2(1-f)}}{3\alpha^2f^2}\bigg)
^{(1+\lambda)}\bigg]^\frac{(m-3)[A+\lambda(1+A)]}{(4-m)(1+A)(1+\lambda)}.
\end{eqnarray}

In order to constraint our model, we must consider the essential condition for warm inflation, $T>H$, the condition for which the model evolves according to the weak regime, $R\ll 1$, and finally the Planck 2015 results \cite{Ade:2015lrj} , through the two-dimensional marginalized joint confidence contours for $n_s$ and $r$, at the 68 and 95 $\%$ CL.  The upper left and upper right plots in Fig.\ref{fig1} show the ratios $T/H$ and $\Gamma/3H$ as functions of the scalar spectral index $n_s$ for the case $m=3$, i.e., $\Gamma (\phi,T)=C_{\phi}T^3/\phi^2$, respectively. To obtain both plots we used three different values for $C_{\phi}$ parameter and considered the following values characterizing the MCG: $A=0\textup{.}0046$, $B=0\textup{.}8289$ (by fixing $\rho_{mcg0}=1$), and $\lambda=0\textup{.}1905$ \cite{Paul:2014kza}, and $C_{\gamma}=70$. In order to obtain numerical values for $T/H$ and $\Gamma/H$, for each value of $C_{\phi}$ we solve numerically the Eqs.(\ref{22}) and (\ref{24}) for $\alpha$ and $f$, considering the observational values
$\mathcal{P}_{\mathcal{R}}\simeq 2\times 10^{-9}$ and $n_s\simeq 0\textup{.}96$ \cite{Ade:2015lrj}, and fixing $N=60$. In this way, for $C_{\phi}=3\times 10^9$, we obtain the values $\alpha=0\textup{.}0062$ and $f=0\textup{.}4703$, whereas for $C_{\phi}=6\times 10^9$, the solution is given by $\alpha=0\textup{.}0043$ and $f=0\textup{.}4702$. Finally, for $C_{\phi}=9\times 10^9$, we find that $\alpha=0\textup{.}0036$ and $f=0\textup{.}4701$. From the upper left panel, we note that for $C_{\phi}>3\times 10^9$, the condition for warm inflation, $T>H$, is always satisfied for all the range considered for $n_s$. On the other hand, from the upper right panel, we note that for $C_{\phi}<9\times 10^9$, the model evolves according to the weak regime, $R\ll 1$. In this way, the condition for warm inflation gives us a lower limit on $C_{\phi}$ and, on the other hand, the condition for which the model evolves in agreement with the weak regime gives us an upper limit for $C_{\phi}$. Then, for the case $m=3$, the allowed range for $C_{\phi}$ become $3\times 10^9<C_{\phi}<9\times 10^9$. 

\begin{figure}[th]
{{\hspace{0cm}\includegraphics[width=3.7in,angle=0,clip=true]{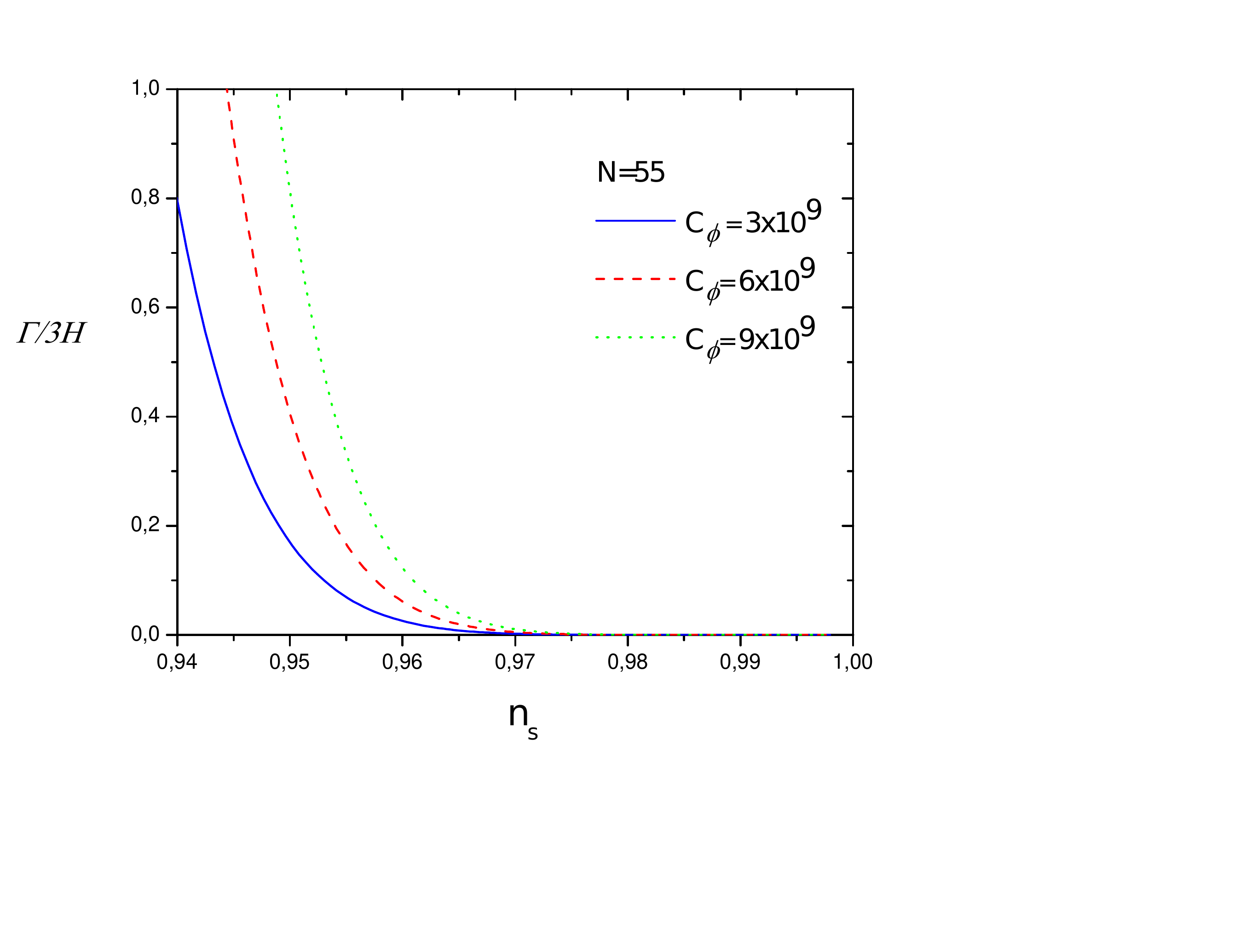}}
\hspace{-2.6 cm}}
{\includegraphics[width=3.7in,angle=0,clip=true]{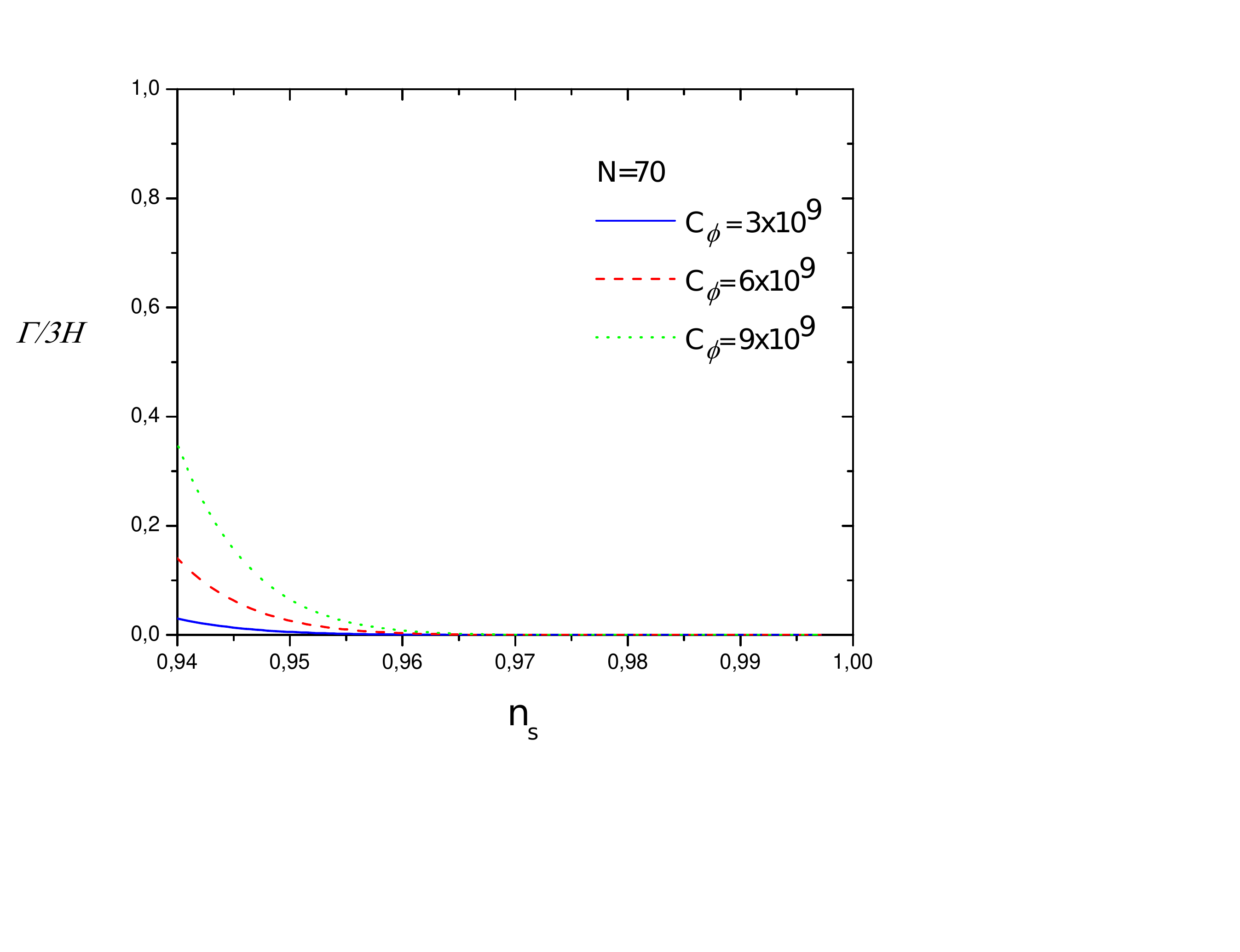}}
{\vspace{-1.5 cm}\caption{ Plots $\Gamma/3H$ as function of the scalar spectral index $n_s$ for $N=55$ and $N=70$, respectively. For the left plot, the dotted, dashed, and
solid lines correspond to the pairs ($\alpha=0\textup{.}0015$, $f=0\textup{.}5237$),
($\alpha=0\textup{.}0011$, $f=0\textup{.}5237$), and ($\alpha=0\textup{.}0008$, $f=0\textup{.}5237$),
respectively. Finally, for the right plot, the dotted, dashed, and
solid lines correspond to the pairs ($\alpha=0\textup{.}5168$, $f=0\textup{.}2531$),
($\alpha=0\textup{.}4704$, $f=0\textup{.}2531$), and ($\alpha=0\textup{.}4456$, $f=0\textup{.}2531$),
respectively. For all plots we have used the values
$C_\gamma=70$, $A=0\textup{.}0046$, $B=0\textup{.}8289$, $\lambda=0\textup{.}1905$, and $\kappa=1$
 \label{fig2}}}
\end{figure}

In addition, to see whether the change on the number of $e$-folds $N$ modifies the allowed range for $C_{\phi}$, firstly we consider $N=55$. We solve numerically Eqs.(\ref{22}) and (\ref{24}) for $\alpha$ and $f$, and considering the observational values
$\mathcal{P}_{\mathcal{R}}\simeq 2\times 10^{-9}$ and $n_s\simeq 0\textup{.}96$ \cite{Ade:2015lrj}. In order to make a direct comparison, with the case $N=60$, we consider the same values already used for $C_{\phi}$. In this way, for $C_{\phi}=3\times 10^9$, we obtain the values $\alpha=0\textup{.}0015$ and $f=0\textup{.}5237$, whereas for $C_{\phi}=6\times 10^9$, the solution is given by $\alpha=0\textup{.}0011$ and $f=0\textup{.}5237$. Finally, for $C_{\phi}=9\times 10^9$, we find that $\alpha=0\textup{.}0008$ and $f=0\textup{.}5237$. Similarly, by fixing $N=70$ and for $C_{\phi}=3\times 10^9$, we obtain the values $\alpha=0\textup{.}5168$ and $f=0\textup{.}2531$, whereas for $C_{\phi}=6\times 10^9$, the solution is given by $\alpha=0\textup{.}4704$ and $f=0\textup{.}2531$. Finally, for $C_{\phi}=9\times 10^9$, we find that $\alpha=0\textup{.}4456$ and $f=0\textup{.}2531$. For $N=55$ and $N=70$, the essential condition for warm inflation to occur, through the plot $T/H$ as function of $n_s$ (not shown) still imposes a lower limit for $C_{\phi}$ which is not modified with respect to $N=60$. However, from left and right panels of Fig.\ref{fig2}, we infer that the condition for the model evolves according to weak regime, modifies the upper limit on $C_{\phi}$ for $N=55$ and $N=70$. In particular, for $N=55$, the new upper limit on $C_{\phi}$ becomes $6\times 10^9$, which is lower than the previous found by fixing $N=60$. However, for $N=70$, the new upper bound becomes $10^{12}$, being greater than the already found for $N=60$. Then, for $N=55$ and $N=70$, the allowed ranges for $C_{\phi}$ are $3\times 10^9<C_{\phi}<6\times 10^9$ and $3\times 10^9<C_{\phi}<10^{12}$, respectively. Having in mind that the changes on $N$ imply a modification on the allowed ranges for $C_{\phi}$, particularly for the upper bound, although not significant, from as now we restrict ourselves to $N=60$.

It is interesting to mention that Planck data, through two-dimensional marginalized joint confidence contours for $n_s$ and $r$, does not impose any constraint on our model for the special case $m=3$. In fact, for the several values considered before, the tensor-to-scalar ratio $r\sim 10^{-6}$ (figure not shown), being compatible with the Planck 2015 data, by considering the two-dimensional marginalized constraints at 68 $\%$ and 95 $\%$ C.L. on the parameters $r$ and $n_s$ \cite{Ade:2015lrj}.

\begin{figure}[th]
{{\hspace{0cm}\includegraphics[width=4in,angle=0,clip=true]{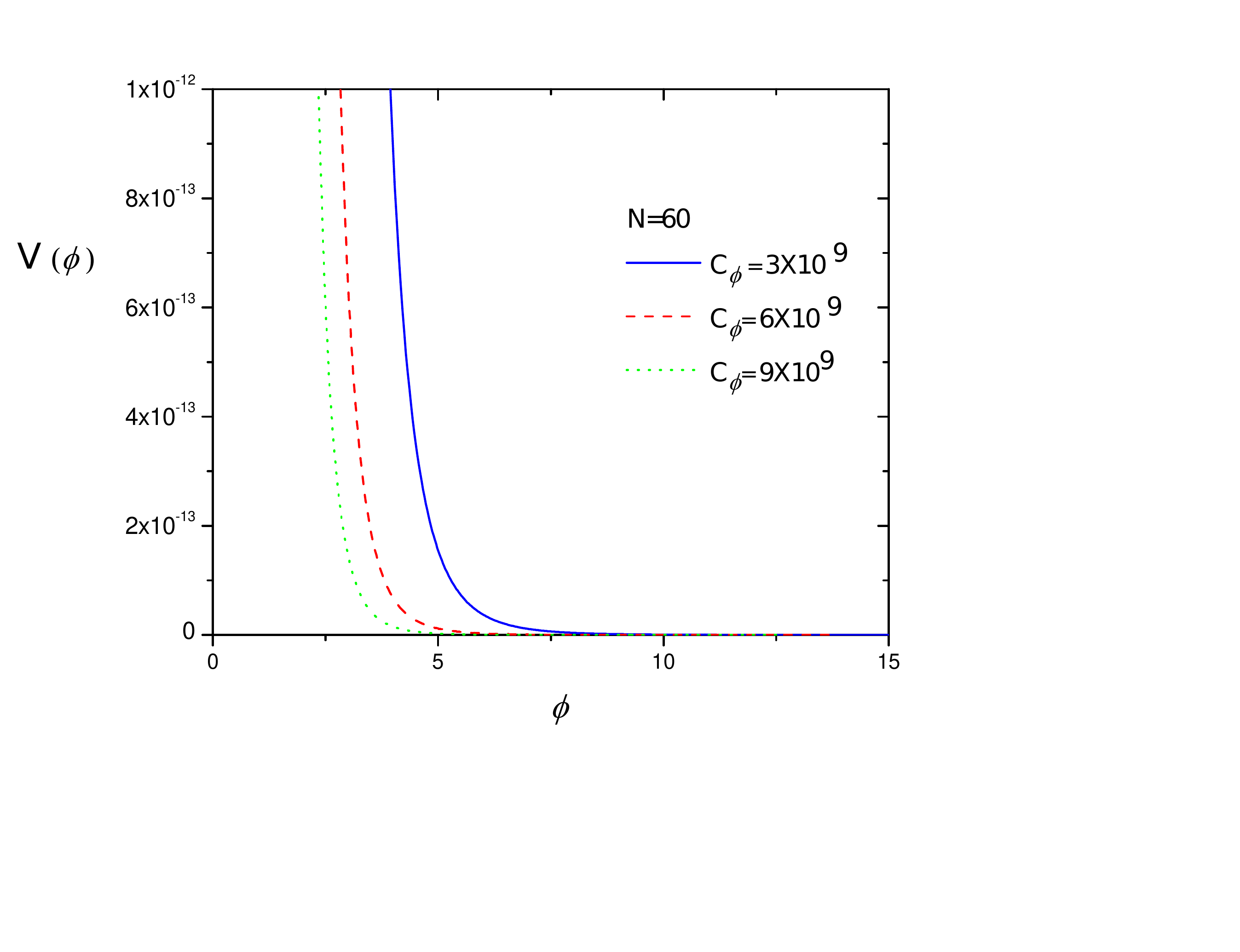}}
}

{\vspace{-1 cm}\caption{ Plot of the effective potential $V$ as function of inflaton field $\phi$ for the case $m=3$ with $N=60$. For this plot we have
considered three different values of the parameter $C_\phi$ for the
special case $m=3$, i.e., $\Gamma\propto T^3/\phi^2$, assuming 
the model evolves according to the weak dissipative regime. The dotted, dashed, and
solid lines correspond to the pairs ($\alpha=0\textup{.}0062$, $f=0\textup{.}4703$),
($\alpha=0\textup{.}0043$, $f=0\textup{.}4702$), and ($\alpha=0\textup{.}0036$, $f=0\textup{.}4701$),
respectively. In addition, we have used the values
$C_\gamma=70$, $A=0\textup{.}0046$, $B=0\textup{.}8289$, $\lambda=0\textup{.}1905$, and $\kappa=1$
 \label{fig3}}}
\end{figure}

{Finally, Fig.\ref{fig3} shows the effective potential, given by Eq.(\ref{13}), as function of the inflaton field $\phi$ in the weak dissipative regime, for the case $m=3$ with $N=60$. Particularly, we have
considered three different values of the parameter $C_\phi$, where the dotted, dashed, and
solid lines correspond to the pairs ($\alpha=0\textup{.}0062$, $f=0\textup{.}4703$),
($\alpha=0\textup{.}0043$, $f=0\textup{.}4702$), and ($\alpha=0\textup{.}0036$, $f=0\textup{.}4701$),
respectively. Inflation takes place as the field rolls down the potential, which tends asymptotically to zero as $\phi\,\rightarrow\,\infty$.

The left and right plots in Fig.\ref{fig4} show the ratios $T/H$ and $\Gamma/3H$ as functions of the scalar spectral index $n_s$ for the case $m=1$, i.e., $\Gamma (\phi,T)=C_{\phi}T$, respectively. To obtain both plots we used three different values for $C_{\phi}$ parameter and considered the same values characterizing the MCG used for the case $m=3$, and $C_{\gamma}=70$. Following the same procedure as for the case $m=3$, for $C_{\phi}=3\times 10^{-3}$ we obtain the values $\alpha=0\textup{.}5168$ and $f=0\textup{.}2531$, whereas for $C_{\phi}=10^{-2}$, the solution is given by $\alpha=0\textup{.}4704$ and $f=0\textup{.}2532$. Finally, for $C_{\phi}=2\times 10^{-2}$, we find that $\alpha=0\textup{.}4456$ and $f=0\textup{.}2533$. From the left panel, we note that for $C_{\phi}>3\times 10^{-3}$, the condition for warm inflation, $\frac{T}{H}>1$, is always satisfied for all the range considered for $n_s$. On the other hand, from the right panel, we note that for $C_{\phi}¡<10^{-2}$, the models evolve according to the weak regime, $R\ll 1$. In this way, the condition for warm inflation gives us a lower limit on $C_{\phi}$ and, on the other hand, the condition for which the model evolves in agreement with the weak regime gives us an upper limit for $C_{\phi}$. Then, for the case $m=1$, the allowed range for $C_{\phi}$ is found to be $3\times 10^{-3}<C_{\phi}<10^{-2}$. Again, the two-dimensional marginalized joint confidence contours for $n_s$ and $r$ dont impose any constraint on $C_{\phi}$. Aditionally, for all the previous values, the tensor-to-scalar ratio $r\sim 10^{-7}$ (figure not shown), supported by Planck 2015 data.

\begin{figure}[th]
{{\hspace{0cm}\includegraphics[width=3.3in,angle=0,clip=true]{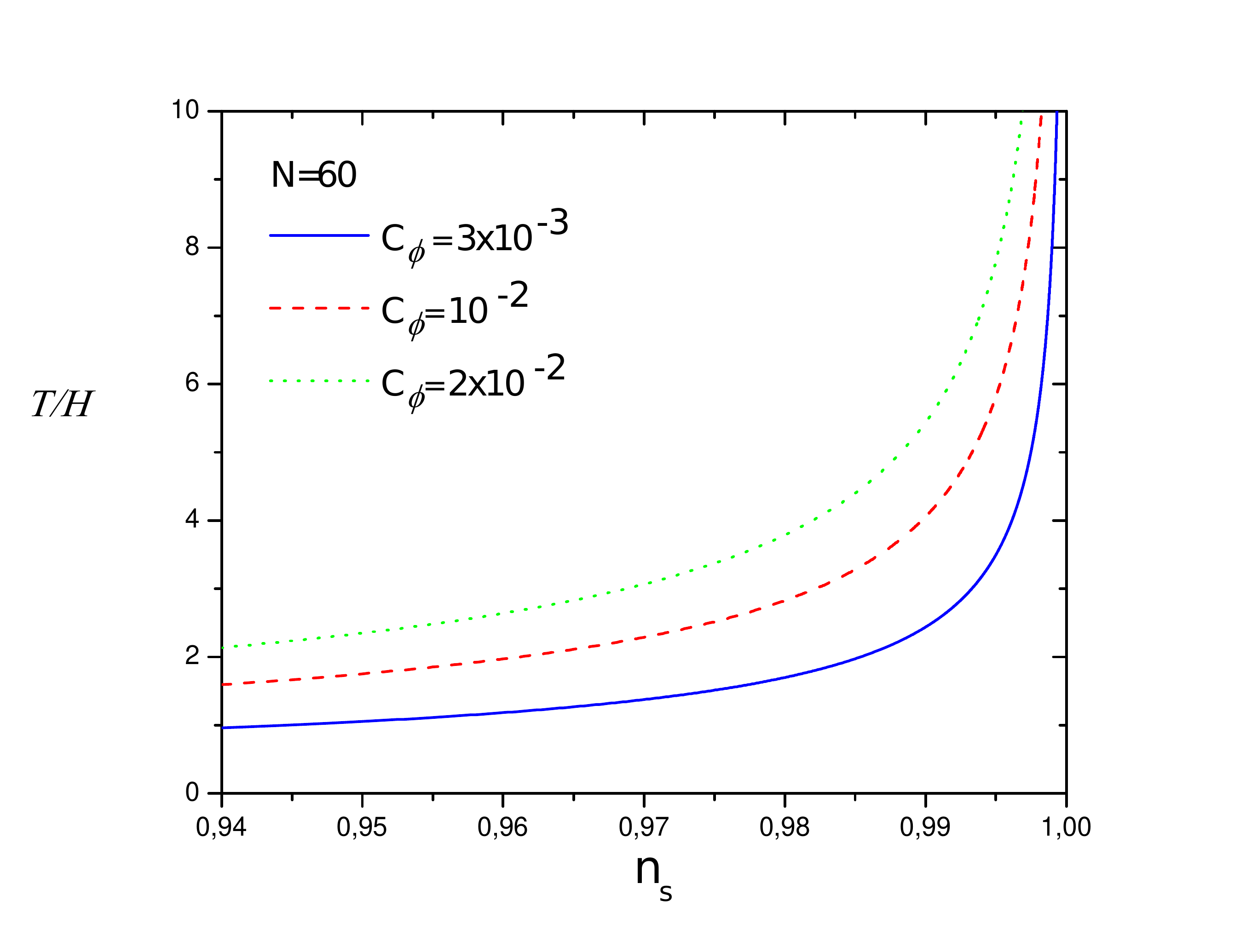}}
\hspace{-0.7 cm}}
{\includegraphics[width=3.3in,angle=0,clip=true]{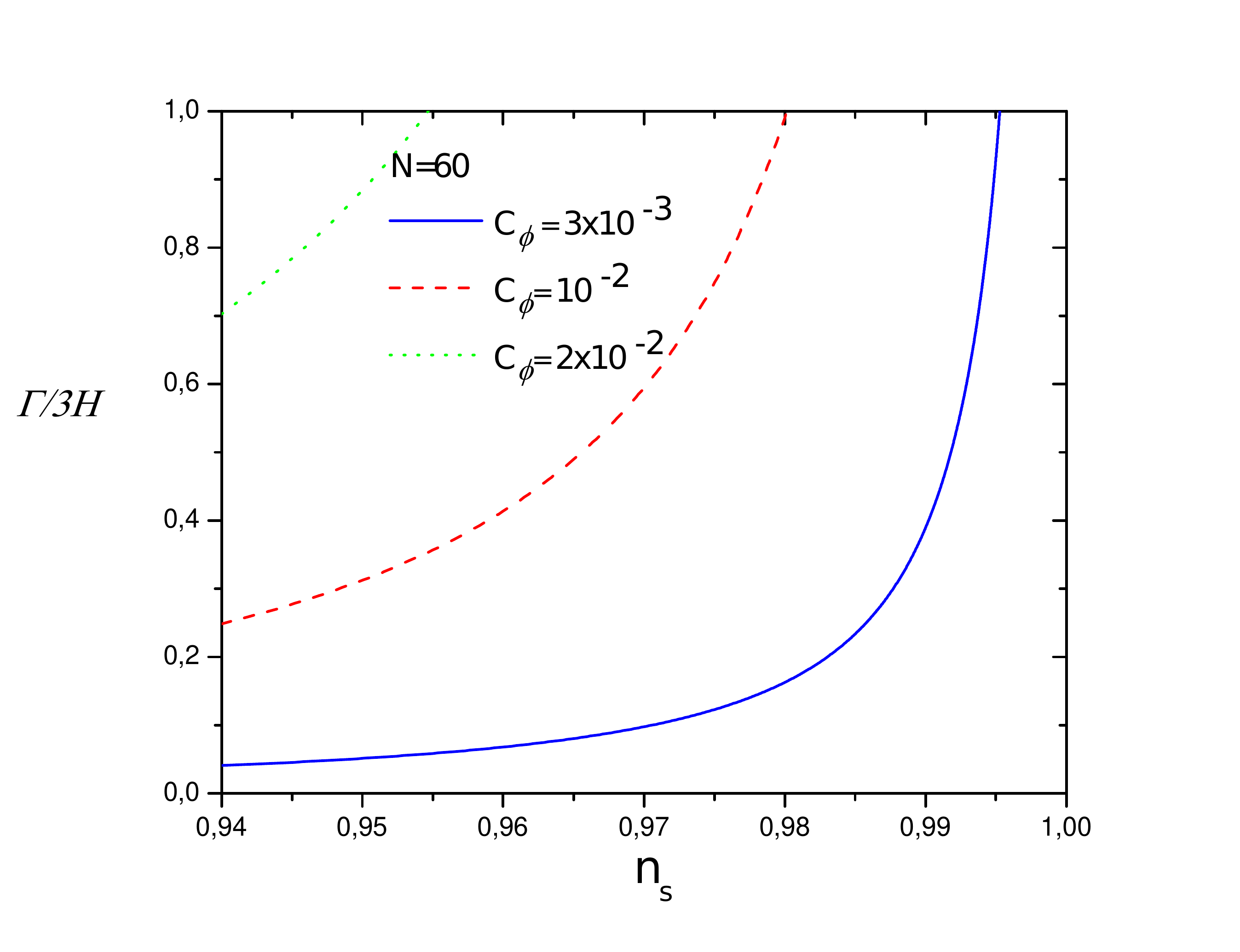}}

{\vspace{-0.5 cm}\caption{ Plots of $T/H$ as function of
the scalar spectral index $n_s$ (left) and $\Gamma/3H$ as function of
the scalar spectral index $n_s$ (right). For both plots we have
considered three different values of the parameter $C_\phi$ for the
special case $m=1$, i.e., $\Gamma\propto T$, assuming 
the model evolves according to the weak dissipative regime. In both panels, the dotted, dashed, and
solid lines correspond to the pairs ($\alpha=0\textup{.}5168$, $f=0\textup{.}2531$),
($\alpha=0\textup{.}4704$, $f=0\textup{.}2532$), and ($\alpha=0\textup{.}4456$, $f=0\textup{.}2533$),
respectively. In these plots we have used the values
$C_\gamma=70$, $A=0\textup{.}0046$, $B=0\textup{.}8289$, $\lambda=0\textup{.}1905$, and $\kappa=1$
 \label{fig4}}}
\end{figure}

As in previous cases, for $m=0$ and $m=-1$, the lower limit on $C_{\phi}$ corresponds to the minimum allowed value for which the essential condition for warm inflation, $\frac{T}{H}>1$, is satisfied, and on the other hand, the upper limit on $C_{\phi}$ correspond to the maximum allowed value for which the model evolves according to the weak regime $R\ll1$. Specifically, for $m=0$, the lower limit
on $C_{\phi}$ is given by $C_{\phi}=1\textup{.}5\times 10^{-9}$, for which we find numerically that $\alpha=0\textup{.}8571$ and $f=0\textup{.}2264$. Additionally, the upper limit on $C_{\phi}$ is found to be $C_{\phi}=1\textup{.}5\times 10^{-8}$. For this value we find numerically $\alpha=0\textup{.}7514$ and $f=0\textup{.}2266$. Finally, for $m=-1$, the lower limit on $C_{\phi}$ corresponds to
$C_{\phi}=5\textup{.}5\times 10^{-16}$, for which we find numerically that $\alpha=1\textup{.}1737$ and $f=0\textup{.}2102$. The upper limit on $C_{\phi}$ is found to be $C_{\phi}=6\textup{.}5\times 10^{-15}$. For this value we find numerically $\alpha=1\textup{.}0458$ and $f=0\textup{.}2104$. Moreover, as in previous cases, we observe that the consistency relation $r(n_s)$ does not impose a constraint on $C_{\phi}$. In this way, for the weak dissipative regime, the constraints on our model are found only by considering the essential condition for warm inflation, $T>H$, and the condition for which the model evolves in agreement with the 
weak dissipative regime, $R\ll 1$.

\section{The Strong Dissipative Regime}\label{strong}

In this section, we analyze the inflationary dynamics of our MCG
model in the strong dissipative regime $\Gamma \gg 3H$. We can find the
solution for the scalar field as function of cosmic time by using Eqs.
(\ref{sc1}) and (\ref{14}). Here, we study the solution for two
cases by separate, for $m=3$ and $m\neq3$.

\subsection{Special Case $m=3$}

For the special case $m=3$, the scalar field as function of cosmic time becomes
\begin{equation}\label{28}
\phi(t)-{\phi}_0=\exp\bigg(\frac{\tilde{M}[t]}{\tilde{S}}\bigg),
\end{equation}
where $\phi(t=0)=\phi_0$ is an integration constant. The quantity
$\tilde{S}$ and the function $\tilde{M}[t]$ are given by
\begin{eqnarray}\nonumber
\tilde{S}&=&2^{-\frac{31}{8}}\frac{{C_{\phi}}^{1/2}}{{C_{\gamma}}
^{3/8}}(1+A)^{-7/8}\frac{(\kappa/3)^{\frac{1}{8(1+A)}}}{({\alpha}f)
^\frac{(3A+5)}{8(1+A)}}(1-f)^{-\frac{1}{8}}\big[(A(4+3f)+5f+2\big].
\\\nonumber \tilde{M}[t]&=&t^\frac{A(4+3f)+5f+2}{8(1+\lambda)}
\textmd{Hypergeometric}2F1\bigg[\frac{A(4+3f)+5f+2}{16(1+A)(1-f)(1+\lambda)},
\\\nonumber&&\frac{A+\lambda(1+A)}{8(1+A)(1+\lambda)},
1+\frac{A(4+3f)+5f+2}{16(1+A)(1-f)(1+\lambda)},
\\\label{29}&&\frac{B}{1+A}\frac{\kappa t^{-2(f-1)(1+\lambda)}}{3{\alpha}^2f^2}\bigg].
\end{eqnarray}
respectively. One can find the Hubble rate for $m=3$ in terms of
scalar field by utilizing Eqs.(\ref{3}) and (\ref{28}) like this
\begin{equation}\label{32}
H(\phi)=\frac{\alpha f}{(\tilde{M}^{-1}[\tilde{S}\ln{\phi}])^{1-f}},
\end{equation}
For this case, the potential $V(\phi)$ leads to
\begin{equation}\label{34}
V(\phi)\approx\bigg[\bigg(\frac{3{\alpha}^2f^2}{\kappa(\tilde{M}
^{-1}[\tilde{S}\ln{\phi}])^{2(1-f)}}\bigg)^{1+\lambda}-\frac{B}{1+A}\bigg]
^{\frac{1}{(1+A)(1+\lambda)}},
\end{equation}
The dissipative coefficient for $m=3$ in terms of scalar field can
be obtained by using Eqs.(\ref{14}) and (\ref{28})
\begin{eqnarray}\nonumber
\Gamma(\phi)&=&{\delta_3}{\phi}^{-2}(\tilde{M}^{-1}[\tilde{S}\ln{\phi}])
^\frac{6A(1-f)-3(2-f)(1+A)}{4(1+A)}
\bigg[1-\frac{B}{1+A}\\\label{36}&\times&\bigg(\frac{\kappa (\tilde{M}^{-1}[\tilde{S}\ln{\phi}])
^{2(1-f)}}{3{\alpha}^2f^2}\bigg)^{1+\lambda}\bigg]
^\frac{-3(A+{\lambda}(1+A))}{4(1+A)(1+\lambda)},
\end{eqnarray}
here $\delta_3$ is a constant and attained the value as
$\delta_3=C_{\phi}\left[\frac{\alpha f(1-f)}{2\kappa C_{\gamma}(1+A)}\right]
^{3/4}(\frac{3\alpha^2f^2}{\kappa})^\frac{-3A}{4(1+A)}$. By combining
Eqs.(\ref{3}) and (\ref{28}), we can find the relation to the number
of $e$-folds $N$ as follows
\begin{equation}\label{38}
N=\int_{t_1}^{t_2}Hdt=\alpha\bigg((\tilde{M}^{-1}[\tilde{S}\ln{\phi}_2])
^f-(\tilde{M}^{-1}[\tilde{S}\ln{\phi}_1])^f\bigg),
\end{equation}

Now, we shall study the cosmological perturbations for our model in the strong dissipative regime
$R=\Gamma/3H>1$. For this regime, the scalar field fluctuation $\delta \phi$is found to be
$\delta{\phi}^2{\simeq}\frac{k_FT}{2{\pi}^2}$ \cite{Berera:2008ar}, where $k_F$ is a
freeze-out wave number which is defined as
$k_F=\sqrt{{\Gamma}H}=H\sqrt{3R}>H$. In this way, the power spectrum of the
scalar perturbation $\mathcal{P}_\mathcal{R}$ can be obtained by
using Eqs.(\ref{12}), (\ref{14}) and (\ref{3}) as
\begin{eqnarray}\nonumber
\mathcal{P}_\mathcal{R}&=&\frac{H^{5/2}{\Gamma}^{1/2}T}{2{\pi}
^2\dot{\phi}^2}=\frac{\kappa (1+A)}{12\pi^2}{C_{\phi}}^{3/2}
\bigg(\frac{3}{\kappa}\bigg)^{-\frac{3A}{8(1+A)}}\bigg[\frac{3}{2\kappa C_{\gamma}(1+A)}\bigg]
^{\frac{11}{8}}{\phi}^{-3}
\\\label{40}&\times&H^{\frac{3(2+A)}{4(1+A)}}(-\dot{H})
^{\frac{3}{8}}\bigg[1-\frac{B}{1+A}\bigg(\frac{3H^2}{\kappa}\bigg)^{-(1+\lambda)}\bigg]
^{-\frac{3(A+\lambda(1+A))}{8(1+A)(1+\lambda)}}.
\end{eqnarray}
In addition, the power spectrum may also be expressed as a function of the scalar field
$\phi$ for $m=3$ by using Eqs.(\ref{3}), (\ref{28}) and (\ref{40})
as
\begin{eqnarray}\nonumber
\mathcal{P}_\mathcal{R}&=&\delta_4(\tilde{M}^{-1}[\tilde{S}\ln{\phi}])
^{\frac{3(f-2)(1+A)-2(1-f)[6(1+A)-3A]}{8(1+A)}}{\phi}^{-3}
\bigg[1-\frac{B}{1+A}\\\label{41}&\times&\bigg(\frac{\kappa (\tilde{M}^{-1}[\tilde{S}\ln{\phi}])
^{2(1-f)}}{3{\alpha}^2f^2}\bigg)^{1+\lambda}\bigg]
^{-\frac{3(A+\lambda(1+A))}{8(1+A)(1+\lambda)}},
\end{eqnarray}
where
$\delta_4=\frac{\kappa (1+A)}{12{\pi}^2}{C_{\phi}}^{3/2}(\frac{3}{\kappa})
^{-\frac{3A}{8(1+A)}}\left[\frac{3}{2kC_{\gamma}(1+A)}\right]^{11/8}(\alpha f)
^{\frac{2[6(1+A)-3A]+3(1+A)}{8(1+A)}}(1-f)^{3/8}$.

Similarly, in terms of the number of $e$-folds $N$, the power spectrum for
$m=3$ becomes
\begin{eqnarray}\nonumber
\mathcal{P}_\mathcal{R}&=&\delta_4(J[N])^{\frac{3(f-2)(1+A)-2(1-f)(6(1+A)-3A)}
{8(1+A)}}\exp\bigg(-\frac{3}{\tilde{S}}\tilde{M}(J[N])\bigg)
\bigg[1-\frac{B}{1+A}\\\label{43}&\times&\bigg(\frac{\kappa (J[N])^{2(1-f)}}{3{\alpha}^2f^2}\bigg)
^{1+\lambda}\bigg]^{\frac{-3(A+\lambda(1+A))}{8(1+A)(1+\lambda)}},
\end{eqnarray}
here, we use Eqs.(\ref{38}) and (\ref{41}). By using Eq.(\ref{41}),
we obtain the scalar spectral index $n_s$ as follow
\begin{equation}\label{45}
n_s=1+\frac{3(f-2)(1+A)-2(1-f)(6(1+A)-3A)}{8\alpha f(1+A)}(\tilde{M}
^{-1}[\tilde{S}\ln{\phi}])^{-f}+n_1+n_2,
\end{equation}
where
\begin{eqnarray}\nonumber
n_1&=&-3\bigg(\frac{6}{\kappa(1+A)}\bigg)^{1/2}
\bigg[\frac{3}{2\kappa C_{\gamma}(1+A)}\bigg]^{-3/8}\bigg(\frac{3{\alpha}^2f^2}{\kappa}\bigg)^{\frac{-A}{8(1+A)}}
(1-f)^{1/8}\\\nonumber&\times&\frac{(\alpha f)^{-3/8}}{{C_\phi}^{1/2}}
\left(\tilde{M}^{-1}[\tilde{S}\ln{\phi}]\right)^{\frac{2A(1-f)-(1+A)(4+3(f-2))}{8(1+A)}}
\bigg[1-\frac{B}{1+A}\\\nonumber&\times&\bigg(\frac{\kappa (\tilde{M}^{-1}[\tilde{S}\ln{\phi}])
^{2(1-f)}}{3{\alpha}^2f^2}\bigg)^{(1+\lambda)}\bigg]^{-\frac{A+\lambda(1+A)}{8(1+\lambda)(1+A)}},
\\\nonumber n_2&=&\frac{3(A+\lambda(1+A))}{4(1+A)}\frac{(\kappa /3)
^{1+\lambda}}{(\alpha f)^{3+2\lambda}}(1-f)(\tilde{M}^{-1}[\tilde{S}\ln{\phi}])
^{2-3f+2\lambda(1-f)}\\\nonumber&\times&\bigg[1-\frac{B}{1+A}\bigg(\frac{\kappa (\tilde{M}^{-1}[\tilde{S}\ln{\phi}])
^{2(1-f)}}{3{\alpha}^2f^2}\bigg)^{1+\lambda}\bigg]^{-1}.
\end{eqnarray}
We can also express the scalar spectral index $n_s$ in terms of
number of $e$-folds $N$ as follows
\begin{equation}\label{47}
n_s=1+\frac{3(f-2)(1+A)-2(1-f)(6(1+A)-3A)}{8\alpha f(1+A)}(J[N])^{-f}+n_1+n_2,
\end{equation}
where $n_1$ and $n_2$ are given by
\begin{eqnarray}\nonumber
n_1&=&-3\bigg(\frac{6}{\kappa (1+A)}\bigg)^{1/2}
\bigg[\frac{3}{2\kappa C_{\gamma}(1+A)}\bigg]^{-3/8}\bigg(\frac{3{\alpha}^2f^2}{\kappa}\bigg)^{\frac{-A}{8(1+A)}}
(1-f)^{1/8}\\\nonumber&\times&\frac{(\alpha f)^{-3/8}}{{C_\phi}^{1/2}}
\bigg(J[N]\bigg)^{\frac{2A(1-f)-(1+A)(4+3(f-2))}{8(1+A)}}
\\\nonumber&\times&\bigg[1-\frac{B}{1+A}\bigg(\frac{\kappa (J[N])
^{2(1-f)}}{3{\alpha}^2f^2}\bigg)^{(1+\lambda)}\bigg]^{-\frac{A+\lambda(1+A)}{8(1+\lambda)(1+A)}}
\\\nonumber
n_2&=&\frac{3(A+\lambda(1+A))}{4(1+A)}\frac{(\kappa /3)^{1+\lambda}}{(\alpha f)
^{3+2\lambda}}(1-f)(J[N])^{2-3f+2\lambda(1-f)}
\bigg[1-\frac{B}{1+A}\\\nonumber&\times&\bigg(\frac{\kappa (J[N])^{2(1-f)}}{3{\alpha}^2f^2}\bigg)^{1+\lambda}\bigg]^{-1}.
\end{eqnarray}
Regarding the tensor perturbations, the tensor-to-scalar ratio in terms of scalar field for $m=3$ leads
to
\begin{eqnarray}\nonumber
r&=&\frac{2\kappa}{\pi^2\mathrm{\delta_4}}(\alpha f)^2(\tilde{M}^{-1}[\tilde{S}\ln{\phi}])
^{\frac{2(1-f)(6(1+A)-3A)-3(f-2)(1+A)-16((1-f)(1+A)}{8(1+A)}}{\phi}^3
\\\label{49}&\times&\bigg[1-\frac{B}{1+A}\bigg(\frac{\kappa (\tilde{M}^{-1}[\tilde{S}\ln{\phi}])
^{2(1-f)}}{3{\alpha}^2f^2}\bigg)^{1+\lambda}\bigg]
^{\frac{3(A+\lambda(1+A))}{8(1+A)(1+\lambda)}}.
\end{eqnarray}
In terms of number of e-folds, the above expression turns out to be
\begin{eqnarray}\nonumber
r&=&\frac{2\kappa}{\pi^2\mathrm{\delta_4}}(\alpha f)^2(J[N])
^{\frac{2(1-f)(6(1+A)-3A)-3(f-2)(1+A)-16((1-f)(1+A)}{8(1+A)}}
\\\label{51}&\times&\exp\bigg({3\frac{\tilde{M}(J[N])}{\tilde{S}}}\bigg)
\bigg[1-\frac{B}{1+A}\bigg(\frac{\kappa (J[N])^{2(1-f)}}{3{\alpha}^2f^2}\bigg)^{1+\lambda}\bigg]
^{\frac{3(A+\lambda(1+A))}{8(1+A)(1+\lambda)}}.
\end{eqnarray}

\begin{figure}[th]
{{\hspace{0cm}\includegraphics[width=3.3in,angle=0,clip=true]{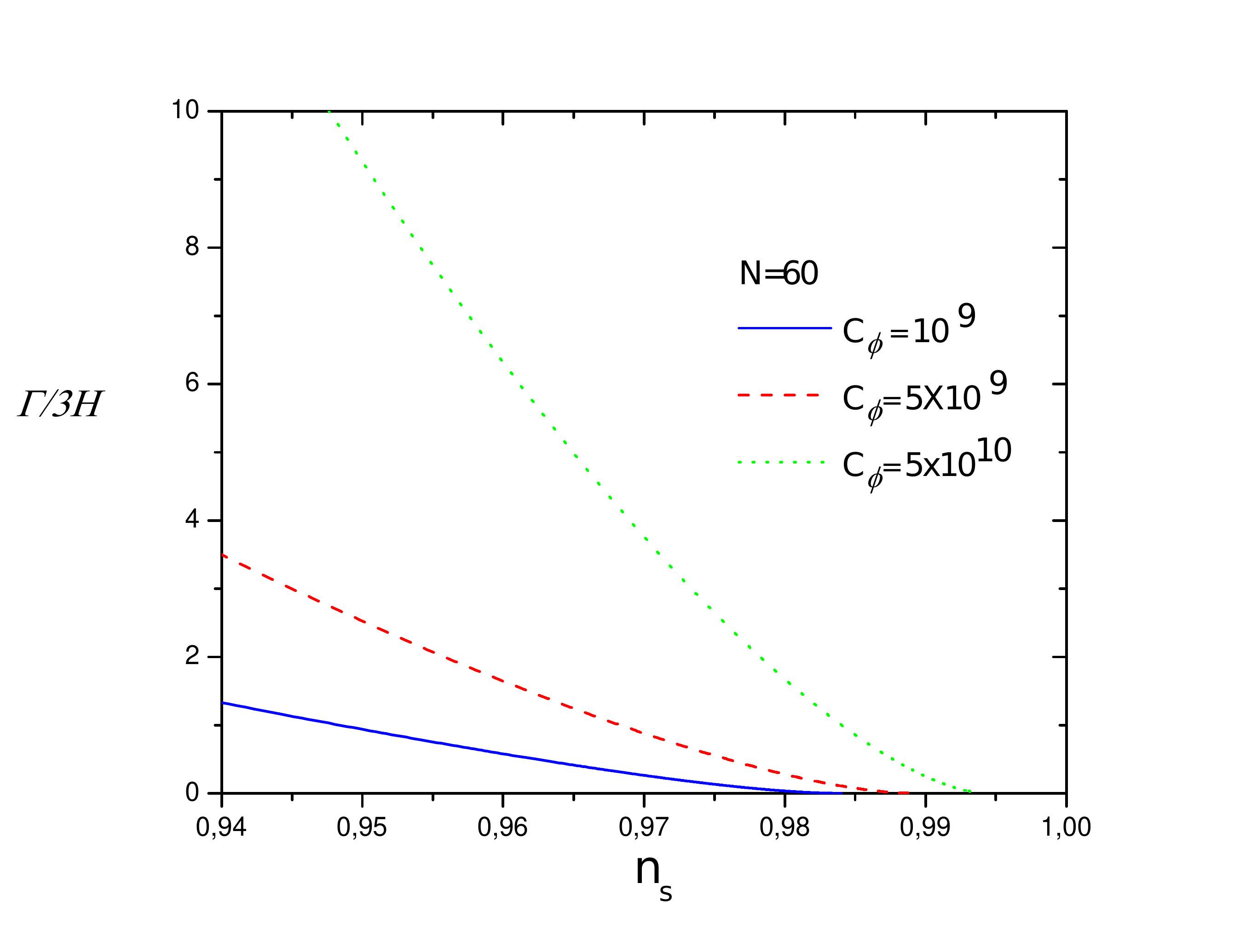}}
\hspace{-0.7 cm}}
{\includegraphics[width=3.3in,angle=0,clip=true]{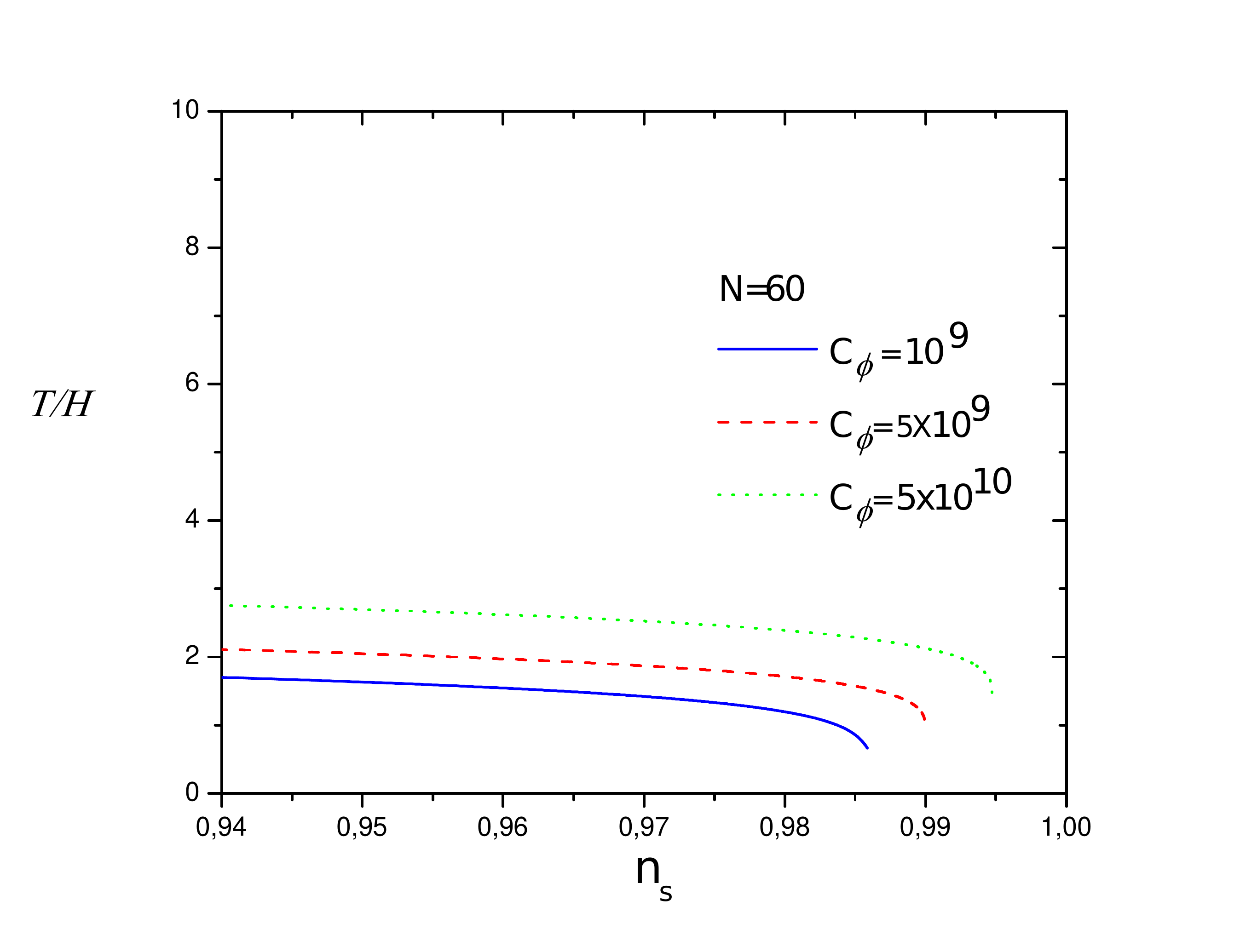}}

{\vspace{-0.5 cm}\caption{ Plots of $\Gamma/3H$ as function of
the scalar spectral index $n_s$ (left) and $T/H$  as function of
the scalar spectral index $n_s$ (right). For both plots we have
considered three different values of the parameter $C_\phi$ for the
special case $m=3$, i.e., $\Gamma\propto T^3/\phi^2$, assuming 
the model evolves according to the strong dissipative regime. In both panels, the dotted, dashed, and
solid lines correspond to the pairs ($\alpha=0\textup{.}0001$, $f=0\textup{.}6451$),
($\alpha=0\textup{.}0003$, $f=0\textup{.}5756$), and ($\alpha=0\textup{.}0004$, $f=0\textup{.}5313$),
respectively. In these plots we have used the values
$C_\gamma=70$, $A=0\textup{.}0046$, $B=0\textup{.}8289$, $\lambda=0\textup{.}1905$, and $\kappa=1$.
 \label{fig5}}}
\end{figure}

In order to constraint our model for this case, in a smiliar way to weak regime, we consider the essential condition for warm inflation, $T>H$, the condition for which the model evolves according to the weak regime, $R\ll 1$, and finally the two-dimensional marginalized joint confidence contours for $n_s$ and $r$, at the 68 and 95 $\%$ CL, by Planck 2015 data \cite{Ade:2015lrj}.
 The left and right plots in Fig.\ref{fig5} show the ratios $\Gamma/3H$ and $T/H$  as functions of the scalar spectral index $n_s$ for the case $m=3$, i.e., $\Gamma (\phi,T)=C_{\phi}T^3/\phi^2$, respectively. To obtain both plots we used three different values for $C_{\phi}$ parameter and the values characterizing the MCG already used: $A=0\textup{.}0046$, $B=0\textup{.}8289$ (by fixing $\rho_{mcg0}=1$), and $\lambda=0\textup{.}1905$ \cite{Paul:2014kza}, and $C_{\gamma}=70$. For each value of $C_{\phi}$ we solve numerically the Eqs.(\ref{22}) and (\ref{24}) for $\alpha$ and $f$, considering the observational values
$\mathcal{P}_{\mathcal{R}}\simeq 2\times 10^{-9}$ and $n_s\simeq 0\textup{.}96$ \cite{Ade:2015lrj}, by fixing $N=60$. In this way, for $C_{\phi}=10^9$, we obtain the values $\alpha=0\textup{.}0001$ and $f=0\textup{.}6451$, whereas for $C_{\phi}=5\times 10^9$, the solution is given by $\alpha=0\textup{.}0003$ and $f=0\textup{.}5756$. Finally, for $C_{\phi}=5\times 10^{10}$, we find that $\alpha=0\textup{.}0004$ and $f=0\textup{.}5313$. From the left panel, we note that for $C_{\phi}>10^9$, the model evolves according to the strong regime, $R\gg 1$. On the other hand, from the right panel, we note that for $C_{\phi}>10^9$ the essential condition for warm inflation, $\frac{T}{H}>1$, is always satisfied. Then, the condition for which the model evolves in agreement with the strong regime gives us an lower limit on $C_{\phi}$. However, the essential condition for warm inflation does not impose any constraint. As sake of comparison, we found numerically that the $\Gamma/3H$ and $T/H$ plots as function of $n_s$ are not not modified when we change the number of $e$-folds to $N=55$ and $N=70$. In this way, the lower limit already found does not change. On the other hand, Fig.\ref{fig6} shows the trajectories in the $n_s-r$ plane along with the two-dimensional marginalized constraints at 68 $\%$ and 95 $\%$ C.L. on the parameters $r$ and $n_s$, by Planck 2015 data  \cite{Ade:2015lrj}. Here, we observe that for $C_{\phi}>5\times 10^9$, the model in the strong dissipative regime is supported by the observational data ($r\sim 10^{-8}$). Then, for the special case $m=3$ with $N=60$, we were able to find only a lower limit for $C_{\phi}$.  

\begin{figure}[th]
{{\hspace{0cm}\includegraphics[width=4.2 in,angle=0,clip=true]{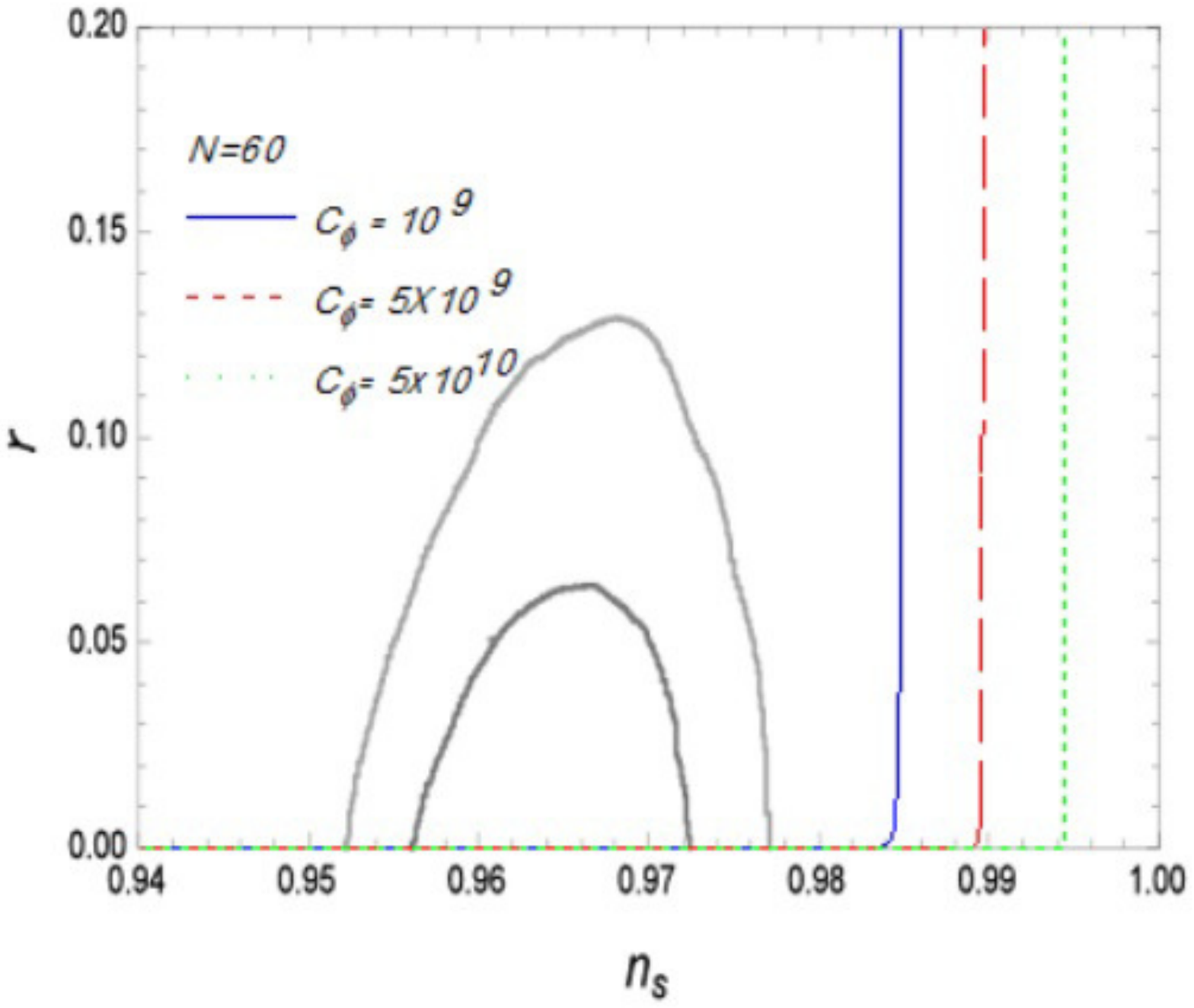}}}

{\vspace{-1 cm}\caption{ Plot of the tensor-to-scalar ratio $r$ versus the scalar spectral index $n_s$ in the strong dissipative regime, for special case $m=3$, i.e., $\Gamma\propto T^3/\phi^2$. In addition, we have considered the two-dimensional marginalized joint confidence contours for $(n_s,r)$, at the 68 and 95 $\%$ C.L., from the latest Planck data \cite{Ade:2015lrj}. In this plot, the dotted, dashed, and solid lines correspond to the pairs ($\alpha=0\textup{.}0001$, $f=0\textup{.}6451$),
($\alpha=0\textup{.}0003$, $f=0\textup{.}5756$), and ($\alpha=0\textup{.}0004$, $f=0\textup{.}5313$),
respectively. Moreover, we have used the values
$C_\gamma=70$, $A=0\textup{.}0046$, $B=0\textup{.}8289$, $\lambda=0\textup{.}1905$, and $\kappa=1$.
 \label{fig6}}}
\end{figure}

\subsection{Special Case $m\neq3$}

The solution for the scalar field for the case $m\neq3$ is found to
be
\begin{equation}\label{30}
\varphi(t)-{\varphi}_0=\frac{\tilde{M}_m[t]}{\tilde{S}_m},
\end{equation}
where $\varphi$ is a new scalar field which is defined as
$\varphi(t)=\frac{2}{3-m}{\phi(t)}^\frac{3-m}{2}$. Also,
$\tilde{S}_m$ and $\tilde{M}_m[t]$ are
\begin{eqnarray}\nonumber
\tilde{S}_m&=&2^{-\frac{(28+m)}{8}}\frac{{C_{\phi}}
^{1/2}}{{C_{\gamma}}^m/8}(1+A)^{-\frac{4+m}{8}}
\frac{(\kappa /3)^{\frac{4-m}{8(1+A)}}}{({\alpha}f)
^\frac{4(1+A)+2A(m-4)-(m-4)(1+A)}{8(1+A)}}
(1-f)^{\frac{m-4}{8}}\\\nonumber&\times&\left[A(4+fm)-f(8-m)+2m-4\right],
\\\nonumber \tilde{M}_m[t]&=&t^\frac{A(4+fm)-f(8-m)+2m-4}{8(1+\lambda)}
\textmd{Hypergeometric}2F1
\\\nonumber&&\bigg[\frac{A(4+fm)-f(8-m)+2m-4}{16(1+A)(1+\lambda)(1-f)},
\frac{(4-m)(A+\lambda(1+A))}{8(1+A)(1+\lambda)},
\\\label{31}&&1+\frac{A(4+fm)-f(8-m)+2m-4}{16(1+A)(1-f)(1+\lambda)},
\frac{B}{1+A}\frac{\kappa t^{-2(f-1)(1+\lambda)}}{3{\alpha}^2f^2}\bigg].
\end{eqnarray}
respectively. Also, in this case, Hubble parameter turns out to be
\begin{equation}\label{33}
H(\varphi)=\frac{\alpha f}{({\tilde{M}_m}^{-1}[\tilde{S}_m{\varphi}])^{1-f}}.
\end{equation}
For this case, the potential $V(\phi)$ takes the form
\begin{equation}\label{35}
V(\varphi)\approx\bigg[\bigg(\frac{3{\alpha}^2f^2}{\kappa({\tilde{M}_m}
^{-1}[{\tilde{S}_m}{\varphi}])^{2(1-f)}}\bigg)^{1+\lambda}-\frac{B}{1+A}\bigg]
^{\frac{1}{(1+A)(1+\lambda)}}.
\end{equation}
Moreover, the dissipative coefficient can be evaluated as
\begin{eqnarray}\nonumber
\Gamma(\phi)&=&{\delta}_m{\phi}^{1-m}({\tilde{M}_m}
^{-1}[\tilde{S}_m{\varphi}])^\frac{2mA(1-f)-m(2-f)(1+A)}{4(1+A)}
\bigg[1-\frac{B}{1+A}\\\label{37}&\times&\bigg(\frac{\kappa ({\tilde{M}_m}
^{-1}[\tilde{S}_m{\varphi}])^{2(1-f)}}{3{\alpha}^2f^2}\bigg)
^{1+\lambda}\bigg]^\frac{-m(A+{\lambda}(1+A))}{4(1+A)(1+\lambda)},
\end{eqnarray}
where
$\delta_m=C_{\phi}\left[\frac{\alpha f(1-f)}{2\kappa C_{\gamma}(1+A)}\right]
^{m/4}\bigg(\frac{3\alpha^2f^2}{\kappa}\bigg)^\frac{-mA}{4(1+A)}$.\\
Also, the number of e-folds become
\begin{equation}\label{39}
N=\alpha\bigg((\tilde{M}_m^{-1}(\tilde{S}_m{\varphi}_2])
^f-(\tilde{M}_m^{-1}[\tilde{S}_m{\varphi}_1])^f\bigg).
\end{equation}
For this case, the power spectrum turns out to be
\begin{eqnarray}\nonumber
\mathcal{P}_\mathcal{R}&=&\tilde{\delta}_m(\tilde{M}_m^{-1}[\tilde{S}_m{\varphi}])
^{\frac{(3m-6)(f-2)(1+A)-2(1-f)[3A(2-m)+6(1+A))]}{8(1+A)}}{\phi}
^{\frac{3(1-m)}{2}}\\\label{42}&\times&\bigg[1-\frac{B}{1+A}\bigg(\frac{\kappa(\tilde{M}_m^{-1}
[\tilde{S}_m{\varphi}])^{2(1-f)}}{3{\alpha}^2f^2}\bigg)^{1+\lambda}\bigg]
^{\frac{(6-3m)[A+\lambda(1+A)]}{8(1+A)(1+\lambda)}},
\end{eqnarray}
where $\tilde{\delta}_m$ is defined as
\begin{eqnarray}\nonumber
\tilde{\delta}_m&=&\frac{\kappa (1+A)}{12{\pi}^2}{C_{\phi}}^{3/2}\bigg(\frac{3}{\kappa}\bigg)
^{\frac{3A(2-m)}{8(1+A)}}\bigg[\frac{3}{2\kappa C_{\gamma}(1+A)}\bigg]
^{\frac{3m+2}{8}}(1-f)^{\frac{3m-6}{8}}\\\nonumber&\times&(\alpha f)
^{\frac{2[3A(2-m)+6(1+A)]+(3m-6)(1+A)}{8(1+A)}}.
\end{eqnarray}
In terms of number of e-folds, we obtain
\begin{eqnarray}\nonumber
\mathcal{P}_\mathcal{R}&=&\tilde{\gamma}_m(J[N])
^{\frac{(3m-6)(f-2)(1+A)-2(1-f)[3A(2-m)+6(1+A))]}{8(1+A)}}(\tilde{M}_m(J[N]))
^{\frac{3(1-m)}{2}}\\\label{44}&\times&\bigg[1-\frac{B}{1+A}\bigg(\frac{\kappa (J[N])
^{2(1-f)}}{3{\alpha}^2f^2}\bigg)^{1+\lambda}\bigg]
^{\frac{(6-3m)(A+\lambda(1+A))}{8(1+A)(1+\lambda)}},
\end{eqnarray}
where $\tilde{\gamma}_m$ is defined as
$\tilde{\gamma}_m=(\frac{1}{\tilde{\delta}_m})^{\frac{3(1-m)}{2}}$.
In this case, the scalar spectrum index $n_s$ becomes
\begin{eqnarray}\nonumber
n_s&=&1+\frac{(3m-6)(f-2)(1+A)-2(1-f)[3A(2-m)+6(1+A)]}
{8\alpha f(1+A)(\tilde{M}_m^{-1}[\tilde{S}_m{\varphi}])^{f}}
\\\label{46}&&+n_{1_m}+n_{2_m},
\end{eqnarray}
where
\begin{eqnarray}\nonumber
n_{1_m}&=&\bigg(\frac{3(1-m)}{2}\bigg)\bigg(\frac{6}{\kappa (1+A)}\bigg)^{1/2}
\bigg[\frac{3}{2\kappa C_{\gamma}(1+A)}\bigg]^{-m/8}(1-f)^{\frac{4-m}{8}}{\phi}^{\frac{m-3}{2}}
\\\nonumber&\times&\bigg(\frac{3{\alpha}^2f^2}{\kappa }\bigg)^{\frac{(m-4)A}{8(1+A)}}
\frac{(\alpha f)^{-m/8}}{{C_\phi}^{1/2}}\bigg(\tilde{M}_m^{-1}[\tilde{S}_m{\varphi}]\bigg)^
{\frac{-2A(1-f)(m-4)-(1+A)(4+m(f-2))}{8(1+A)}}\\\nonumber&\times&
\bigg[1-\frac{B}{1+A}\bigg(\frac{\kappa (\tilde{M}_m^{-1}[\tilde{S}_m{\varphi}])^{2(1-f)}}{3{\alpha}^2f^2}\bigg)^{(1+\lambda)}\bigg]
^{\frac{(m-4)(A+\lambda(1+A))}{8(1+\lambda)(1+A)}}, \\\nonumber
n_{2_m}&=&-\bigg(\frac{6-3m}{4}\bigg)\bigg(\frac{A+\lambda(1+A)}{1+A}\bigg)\frac{(\kappa /3)
^{1+\lambda}(1-f)}{(\alpha f)^{3+2\lambda}}(\tilde{M}_m^{-1}[\tilde{S}_m{\varphi}])
^{2-3f+2\lambda(1-f)}\\\nonumber&\times&\bigg[1-\frac{B}{1+A}\bigg(\frac{\kappa (\tilde{M}_m^{-1}[\tilde{S}_m{\varphi}])
^{2(1-f)}}{3{\alpha}^2f^2}\bigg)^{1+\lambda}\bigg]^{-1}.
\end{eqnarray}
The scalar spectral index in terms of $N$ becomes
\begin{eqnarray}\nonumber
n_s&=&1+\frac{(3m-6)(f-2)(1+A)-2(1-f)[3A(2-m)+6(1+A)]}{8\alpha f(1+A)(J[N])^{f}}
\\\label{48}&&+n_{1_m}+n_{2_m},
\end{eqnarray}
where
\begin{eqnarray}\nonumber
n_{1_m}&=&\bigg(\frac{3(1-m)}{2}\bigg)\bigg(\frac{6}{\kappa (1+A)}\bigg)^{1/2}
\bigg[\frac{3}{2\kappa C_{\gamma}(1+A)}\bigg]^{-m/8}(1-f)^{\frac{4-m}{8}}{\phi}^{\frac{m-3}{2}}
\\\nonumber&\times&\bigg(\frac{3{\alpha}^2f^2}{\kappa }\bigg)^{\frac{(m-4)A}{8(1+A)}}
\frac{(\alpha f)^{-m/8}}{{C_\phi}^{1/2}}(J[N])^
{\frac{-2A(1-f)(m-4)-(1+A)(4+m(f-2))}{8(1+A)}}\\\nonumber&\times&
\bigg[1-\frac{B}{1+A}\bigg(\frac{\kappa (J[N])^{2(1-f)}}{3{\alpha}^2f^2}\bigg)^{(1+\lambda)}\bigg]
^{\frac{(m-4)(A+\lambda(1+A))}{8(1+\lambda)(1+A)}} \\\nonumber
n_{2_m}&=&-\frac{(6-3m)[A+\lambda(1+A)]}{4(1+A)}\frac{(\kappa /3)^{1+\lambda}}{(\alpha f)
^{3+2\lambda}}(1-f)(J[N])^{2-3f+2\lambda(1-f)}\\\nonumber&\times&
\bigg[1-\frac{B}{1+A}\big(\frac{\kappa (J[N])^{2(1-f)}}{3{\alpha}^2f^2}\bigg)^{1+\lambda}\bigg]^{-1}.
\end{eqnarray}

The tensor-to-scalar ratio takes the following form
\begin{eqnarray}\nonumber
r&=&\frac{2\kappa }{\pi^2\tilde{\delta}_m}(\alpha f)^2(\tilde{M}_m^{-1}[\tilde{S}_m{\varphi}])
^{\frac{2(1-f)[3A(2-m)+6(1+A)]-(3m-6)(f-2)(1+A)-16((1-f)(1+A)}{8(1+A)}}
\\\label{50}&\times&{\phi}
^{\frac{3(m-1)}{2}}\bigg[1-\frac{B}{1+A}\bigg(\frac{\kappa (\tilde{M}_m^{-1}[\tilde{S}_m{\varphi}])
^{2(1-f)}}{3{\alpha}^2f^2}\bigg)^{1+\lambda}\bigg]
^{\frac{(3m-6)(A+\lambda(1+A))}{8(1+A)(1+\lambda)}},
\end{eqnarray}
in terms of the number of $e$-folds
\begin{eqnarray}\nonumber
r&=&\frac{2\kappa}{\pi^2\tilde{\delta}_m}(\alpha f)^2(J[N])
^{\frac{2(1-f)[3A(2-m)+6(1+A)]-(3m-6)(f-2)(1+A)-16((1-f)(1+A)}{8(1+A)}}
\\\label{52}&\times&(\tilde{M}_m(J[N]))
^{\frac{3(m-1)}{2}}\bigg[1-\frac{B}{1+A}\bigg(\frac{k(J[N])^{2(1-f)}}{3{\alpha}^2f^2}\bigg)
^{1+\lambda}\bigg]^{\frac{(3m-6)(A+\lambda(1+A))}{8(1+A)(1+\lambda)}}.
\end{eqnarray}

For the case $m=1$, the condition for the model evolves acording to strong dissipative regime, $R\gg 1$, gives us the lower limit on
$C_{\phi}$, yielding $C_{\phi}=6\times 10^{-2}$ (plot not shown). Additionally, for $C_{\phi}>6\times 10^{-2}$ the condition for warm inflation, $\frac{T}{H}>1$, is always satisfied. Then, we can not find an upper limit on $C_{\phi}$ by considering the $T/H$ plot. Moreover, for $C_{\phi}>6\times 10^{-2}$, the tensor-to-scalar ratio becomes $r\sim 0$, but the model is still supported by the last data of Planck, by considering the two-dimensional marginalized joint confidence contours for $(n_s,r)$, at the 68 and 95 $\%$ C.L. (plot not shown). Then, for the case $m=1$, we were only able to find a lower limit for $C_{\phi}$, given by $C_{\phi}=6\times 10^{-2}$.

For $m=-1$ and $m=0$, the predicted scalar spectral index is always greater than unity, being discarded  by obervations. This means that the inflaton decay ratios $\Gamma \propto \phi$ and $\Gamma \propto \frac{\phi^2}{T}$ are not suitable to describe a strong dissipative dynamics in the MCG scenario. It interesting to mention that same beheaviour has been already reported in \cite{35H, 18}.

\section{Conclusions}\label{conclu}

In the present work we have studied the warm inflationary 
dynamics inspired by the modified Chaplygin gas. We considered
the inflationary expansion was driven by a standard scalar field with
a generalized expression for its decay ratio $\Gamma=C_{\phi}T^m/{\phi}^{m-1}$,
where $m=3,1,0,-1$, denotes several inflaton decay ratios studied in the literature.
We have solved the background as well as perturbative dynamics considering 
the model evolves according to (i) strong and (ii) strong disipative regimes. 
For each dissipative regime, under the slow-roll approximation, we 
have found the expressions for the  scalar power spectrum, scalar spectral index and tensor-to-scalar ratio
subsequently. Contrary to the standard cold inflation, in the warm inflation scenario it is not
sufficient to consider only the constraints on the $r$-$n_s$ plane, but we also have to consider
the essential condition for warm inflation $T>H$ and the conditions for the model evolves under the weak ($R\ll 1$) or strong  ($R\gg 1$)
dissipative regimes. In partcular, for the weak disipative regime, the condition for warm inflation and the condition for the model evolves according to this regime, set the lower and upper limit for the disipative parameter $C_{\phi}$, respectively. The Planck data, by considering the two-dimensional marginalized constraints at 68 $\%$ and 95 $\%$ C.L. on the parameters $r$ and $n_s$, does not impose any constraints on the model for this dissipative regime. However, the values for tensor-to-scalar ratio $r$ are compatible with
current observational data. Regarding the strong dissipative, for the special case $m=3$, the condition for the model evolves under this regime and the Planck data, through the two-dimensional marginalized constraints on the parameters $r$ and $n_s$ set the lower and upper limits on the dissipative parameter $C_{\phi}$. However, for the case $m=1$, neither the condition for warm inflation nor the two-dimensional marginalized constraints on the parameters $r$ and $n_s$ impose contraints on $C_{\phi}$. The condition for the model evolves under the strong regime only sets a lower limit for this quantity. Finally, the both cases $m=0$ and $m=-1$ fail in describe a strong dissipative dynamics consistent with current data, since the predicted value for the scalar spectral index is always greater that unity. It is interesting to mention that the inflationary dynamics of our model under the strong regime predicts a value for the tensor-to-scalar ratio $r\sim 0$, but compatible with current data. We conclude that  warm intermediate inflation inspired by modified Chaplygin gas is compatible with current data for all the several inflaton decay ratios, parametrized by $m$, if we assume that our model evolves under the dissipative regime. However, if we assume that our model takes place in the strong dissipative regime, only the inflaton decay ratios yielding a dynamics compatible with current data correspond to $m=3$ and $m=1$.

\begin{acknowledgments}
N.V. was supported by Comisi\'on Nacional
de Ciencias y Tecnolog\'ia of Chile through FONDECYT Grant N$^{\textup{o}}$
3150490. Finally, the authors wish
to thank the anonymous referee for her/his valuable comments, which
have helped us to improve the presentation in our manuscript.
\end{acknowledgments}

\end{document}